\documentstyle[times,pramana,epsf,floats]{ias}
\def\la{\left\langle} \def\ra{\right\rangle}
\begin{document} 
\title{Statistical Properties of Turbulence: An Overview}

\author{Rahul Pandit$^1$, Prasad Perlekar, and Samriddhi Sankar Ray}
\address{Centre for Condensed Matter Theory, \\ Department of 
Physics,\\
Indian Institute of Science, \\ Bangalore 560012,\\India.  \\$^1$ Also at :
JNCASR, Bangalore, India} 
\abstract{We present an introductory overview of several challenging problems in the statistical characterisation of turbulence. We provide examples from
fluid turbulence in three and two dimensions, from the turbulent advection of passive scalars,  turbulence in the one-dimensional
Burgers equation, and fluid turbulence in the presence of polymer additives.} 
\maketitle

\section{Introduction} 

Turbulence is often described as the last great unsolved problem of
classical  physics~\cite{ecke,falcosreeni,procsreeni}.  However, it is not
easy to state what would constitute a solution of the turbulence problem.
This is  principally because turbulence is not {\it one problem} but a
collection of  {\it several} important problems: These include the
characterisation and  control of turbulent flows, both subsonic and
supersonic, of interest to engineers such as flows in pipes or over cars and
aeroplanes~\cite{engg,pope}.  Mathematical questions in this area are
concerned with developing proofs of the smoothness, or lack thereof, of
solutions of the Navier-Stokes and related
equations~\cite{math,clay,const,doe95,majdabert}.  Turbulence also provides
a variety of challenges for fluid
dynamicists~\cite{pope,tenlum,lesieur,davidson},
astrophysicists~\cite{astro,arnab,krishan,goedbloed},
geophysicists~\cite{geo,boe83}, climate scientists~\cite{climate}, plasma
physicists~\cite{arnab,krishan,goedbloed,biskamp,mkvrev}, and statistical
physicists~\cite{orszag,rosesulem,dommartin,yakors,mccomb,eyink,bohr,adzhem,jkbbook,nazarenko}.
In this brief overview, written primarily for physicists who are not experts
in turbulence, we concentrate on some recent advances in the statistical
characterisation of fluid turbulence~\cite{frisch} in three dimensions, the
turbulence of passive scalars such as  pollutants~\cite{falcormp},
two-dimensional turbulence in thin films or soap
films~\cite{tabelingrev,goldburgrev}, turbulence in the Burgers
equation~\cite{burgers,fri00,bec07}, and fluid turbulence with polymer
additives~\cite{lumleyrev,procdragrev,drag08}; in most of this paper we
restrict ourselves to {\it homogeneous, isotropic turbulence}
~\cite{frisch,batchelor,moninyaglom}; and we highlight some similarities
between the statistical properties of systems at a critical point and those
of turbulent fluids~\cite{jkbbook,epjb,ssrnjp}.  Several important problems
that we do not attempt to  cover include Rayleigh-B\'enard
turbulence~\cite{lohse}, superfluid turbulence~\cite{procsreeni,superfluid},
magnetohydrodyanmic turbulence~\cite{arnab,goedbloed,biskamp,mkvrev}, the
behaviour of inertial particles in turbulent flows~\cite{inertial}, the
transition to turbulence in different experimental
situations~\cite{drazin,bruno}, and boundary-layer~\cite{schlichting,nagib}
and wall-bounded~\cite{walls} turbulence.

This paper is organised as follows: Section 2 gives an overview of some of
the experiments of relevance to our discussion here.  In Section 3 we
introduce the equations that we consider.  Section 4 is devoted to a
summary of phenomenological approaches that have been developed, since the
pioneering studies of Richardson~\cite{rich} and Kolmogorov~\cite{K41}, in
1941 (K41), to understand the behaviour of velocity and other structure
functions in {\it inertial ranges}.  Section 5 introduces the ideas of
multiscaling that have been developed to understand deviations from the
predictions of K41-type phenomenology.  Section 6 contains illustrative
direct numerical simulations; it consists of five subsections devoted to
(a) three-dimensional fluid turbulence, (b) shell models, (c) two-dimensional turbulence in
soap films, (d) turbulence in the one-dimensional Burgers equation, and
(e) fluid turbulence with polymer additives. Section 7 contains concluding
remarks.

\section{Experimental Overview}

Turbulent flows abound in nature. They include the flow of water in a 
garden pipe or in rapids, the flow of air over moving cars or aeroplanes, 
jets that are formed when a fluid is forced through an orifice, the 
turbulent advection of pollutants such as ash from a volcanic eruption, 
terrestrial and Jovian storms, turbulent convection in the sun, and 
turbulent shear flows in the arms of spiral galaxies. A wide variety of 
experimental studies have been carried out to understand the properties 
of such turbulent flows; we concentrate on those that are designed to 
elucidate the statistical properties of turbulence, especially turbulence 
that is, at small spatial scales and far away from boundaries, {\it 
homogeneous and isotropic}. Most of our discussion will be devoted 
to incompressible flows, i.e., low-Mach-number cases in which the fluid 
velocity is much less than the velocity of sound in the fluid. 

In laboratories such turbulence is generated in many different ways. A
common method uses a grid in a wind tunnel~\cite{grid}; the flow
downstream from this grid is homogeneous and isotropic, to a good
approximation. Another technique use the von K\'arm\'an swirling flow,
i.e., flow generated in a fluid contained in a cylindrical tank with two
coaxial, counterrotating discs at its
ends~\cite{pintonacc,bodenacc,ictr1}; in the middle of the tank, far away
from the discs, the turbulent flow is approximately homogeneous and
isotropic.  Electromagnetically forced thin films and soap
films~\cite{ecke,tabelingrev,goldburgrev} have yielded very useful results
for two-dimensional turbulence.  Turbulence data can also be obtained from
atmospheric boundary
layers~\cite{atmosblayer1,atmosblayer2,tsinober,ciclope}, oceanic
flows~\cite{ocean}, and astrophysical measurements~\cite{astro};
experimental conditions cannot be controlled as carefully in such natural
settings as they can be in a laboratory, but a far greater range of length
scales can be probed than is possible in laboratory experiments.

Traditionally, experiments have measured the velocity ${\bf u}({\bf x},t)$
at a single point ${\bf x}$ at various times $t$ by using hot-wire
anemometers; these anemometers can have limitations in (a) the number of
components of the velocity that can be measured and (b) the spatial and
temporal resolutions that can be obtained~\cite{measurements,dpiv}.  Such
measurements yield a time series for the velocity; if the mean flow
velocity $U >> u_{rms}$, the root-mean-square fluctuations of the velocity,
then Taylor's frozen-flow hypothesis~\cite{pope,frisch} can be used to
relate temporal separations $\delta t$ to spatial separations $\delta r$,
along the mean flow direction via $\delta r = U \delta t$. The Reynolds
number $Re = U L/\nu$, where $U$ and $L$ are typical velocity and length
scales in the flow and $\nu$ is the kinematic viscosity, is a convenient
dimensionless control parameter; at low $Re$ flows are laminar; as it
increases increases there is a transition to turbulence often via a variety
of instabilities~\cite{drazin} that we will not cover here; and at large
$Re$ fully developed turbulence sets in. To compare different flows it is
often useful to employ the Taylor-microscale Reynolds number  $Re_\lambda =
u_{rms} \lambda/\nu$, where the Taylor microscale  $\lambda$ can be obtained
from the energy spectrum as described below  (Sec. 6.3).

Refinements in hot-wire anemometry~\cite{tsinober,multihwa} and flow 
visualisation techniques such as laser-doppler velocimetry 
(LDV)~\cite{measurements}, particle-image velocimetry 
(PIV)~\cite{measurements,dpiv}, particle-tracking velocimetry
(PTV)~\cite{measurements,dpiv}, tomographic PIV~\cite{tompiv},
holographic PIV~\cite{holpiv}, and digital holographic
microscopy~\cite{digholmic} have made it possible to obtain reliable
measurements of the Eulerian velocity ${\bf u}({\bf x},t)$ (see Sec. 3) 
in a turbulent flow. In the simplest forms of anemometry a time series of 
the velocity is obtained at a given point in space; in PIV two components
of the velocity field can be obtained in a sheet at a given time; 
holographic PIV can yield all components of the velocity field in a 
volume. Components of the  velocity derivative tensor $A_{ij} \equiv 
\partial_j u_i$ can also be  obtained~\cite{tsinober} and thence quantities 
such as the energy dissipation rate per unit mass per unit volume 
$\epsilon \equiv -\nu \sum_{i,j} (\partial_i u_j + \partial_j u_i)^2$, the 
vorticity $\omega = \nabla  \times {\bf  u}$, and components of the rate 
of strain tensor $s_{ij} \equiv (\partial_i u_j + \partial_j u_i)/2$, where 
the subscripts $i$ and $j$ are Cartesian indices. A discussion of the 
subtleties and limitations of these measurement techniques lies beyond 
the scope of our overview; we refer the reader to 
Refs.~\cite{tsinober,measurements,dpiv} for details. 
Significant progress has also been made over the past decade in the 
measurement of Lagrangian trajectories (see Sec. 3) of tracer particles in 
turbulent flows~\cite{pintonacc,bodenacc}. Given such measurements, 
experimentalists can obtain several properties of turbulent flows. We 
give illustrative examples of the types of  properties we consider.

Flow-visualisation methods often display large-scale coherent structures
in turbulent flows.  Examples of such structures
plumes in Rayleigh-B\'ernard
convection~\cite{vandyke}, structures behind a splitter
plate~\cite{roshko}, and large vortical structures in two-dimensional or
stratified flows~\cite{ecke,tabelingrev,goldburgrev}.  In
three-dimensional flows, as we will see in greater detail below, energy
that is pumped into the flow at the injection scale $L$ cascades, as first
suggested by Richardson~\cite{rich}, from large-scale eddies
to small-scale ones till it is eventually dissipated around and beyond the
dissipation scale $\eta_d$. By contrast, two-dimensional
turbulence~\cite{tabelingrev,goldburgrev,kraichmont,kraic67} 
displays a dual
cascade: there is an inverse  cascade of energy from the scale at which it
is  pumped into the system to large length scales and a direct cascade of
enstrophy $\Omega = \langle \frac{1}{2}\omega^2 \rangle$ to small length
scales. The inverse cascade of energy is associated with the formation of
a few large vortices; in practical realisations the sizes of such vortices
are controlled finally by Ekman friction that is induced, e.g., by air
drag in soap-film turbulence.

Measurements of the vorticity $\omega$ in highly turbulent flows show that
regions of large $\omega$ are organised  into slender tubes. The first
experimental evidence for this was obtained by seeding the flow with
bubbles that moved preferentially to regions of low
pressure~\cite{tubes91} that are associated with large-$\omega$ regimes.
For recent experiments on vortex tubes we refer the reader to
Ref.~\cite{tubes09}.

The time series of the fluid velocity at a given point ${\bf x}$ shows
strong fluctuations. It is natural, therefore, to inquire into the
statistical properties of turbulent flows.  From the Eulerian velocity ${\bf
u}({\bf x},t)$ and its derivatives we can obtain one-point statistics, such
as probability distribution functions (PDFs) of the velocity and its
derivatives. Velocity PDFs are found to be close to Gaussian distributions.
However, PDFs of $\omega^2$ and velocity derivatives show significant
non-Gaussian tails; for a recent study, which contains  references to
earlier work, see Ref.~\cite{tsinober}. The PDF of $\epsilon$ is
non-Gaussian too and the time series of $\epsilon$ is highly
intermittent~\cite{mensreeni}; furthermore, in the limit $Re \to \infty$,
i.e., $\nu \to 0$, the energy dissipation rate per unit volume $\epsilon$
approaches a positive constant value (see, e.g., Fig. 2 of
Ref.~\cite{kanedarev}), a result referred to as a {\it dissipative anomaly}
or the {\it zeroth law of turbulence}.  

Various statistical properties of the rate-of-strain tensor, with components
$s_{ij}$, have been measured~\cite{tsinober}.  The eigenvalues $\lambda_1,
\, \lambda_2,$ and $\lambda_3$, with $\lambda_1 > \lambda_2 > \lambda_3$ ,
of this  tensor must satisfy $\lambda_1 + \lambda_2 + \lambda_3 = 0$, with
$\lambda_1 > 0$ and $\lambda_2 < 0$, in an incompressible flow. The sign of
$\lambda_2$ cannot be determined by this condition but its PDF shows that,
in turbulent flows, $\lambda_2$ has a small, positive mean
value~\cite{chirag02}; and the PDFs of $\cos(\omega\cdot e_i)$, where $e_i$
is the normalised eigenvector corresponding to $\lambda_i$, show that there
is a preferential alignment~\cite{tsinober} of $\omega$ and $e_2$.  Joint
PDFs can be measured too with good accuracy. An example of recent interest
is a tear-drop feature observed in contour plots of the joint PDF of,
respectively, the
second and third invariants, $Q=-tr(A^2)/2$ and $R=-tr(A^3)/3$ 
of the velocity gradient tensor $A_{ij}$ (see Fig. 11 of Ref.~\cite{tsinober}); we display such a
plot in Sec. 6 that deals with direct numerical simulations. 

Two-point statistics are characterised conventionally by studying the
equal-time, order-$p$, longitudinal velocity structure function
\begin{equation} S_p({\bf r}) = \langle [(u({\bf x} + {\bf r}) - u({\bf
x}))\cdot({\bf r}/r]^p\rangle, \end{equation} where the angular brackets
indicate a time average over the nonequilibrium statistical steady state
that we obtain in forced turbulence (decaying turbulence is discussed in
Sec. 6.2). Experiments~\cite{frisch,gagne} show that, for separations $r$
in the {\it inertial range} $\eta_d << r << L$, 
\begin{equation} 
S_p({\bf r}) \sim r^{\zeta_p}, 
\end{equation} 
with exponents $\zeta_p$ that deviate
significantly from the simple scaling prediction~\cite{K41} $\zeta_p^{K41}
= p/3$, especially for $p > 3$, where $\zeta_p < \zeta_p^{K41}$. This
prediction, made by Kolmogorov in 1941 (hence the abbreviation K41), is
discussed in Sec. 4 below; the deviations from this simple scaling
prediction are referred to as multiscaling (Sec. 5) and they are
associated with the intermittency of $\epsilon$ mentioned above. 
We mention, in passing, that the log-Poisson model due to She and 
Leveque provides a good parametrisation of the plot of $\zeta_p$ 
versus $p$ ~\cite{she}.

The second-order structure function $S_2({\bf r})$ can be related easily by
Fourier transformation to the the energy spectrum $E(k) = 4\pi
k^2\langle|{\tilde {\bf u}}(k)|^2\rangle$, where the tilde denotes the
Fourier transform, $k = |{\bf k}|$, ${\bf k}$ is the wave vector, we assume
that the turbulence is homogeneous and isotropic, and, for specificity, we
give the formula for the three-dimensional case.  Since $\zeta_2^{K41} =
2/3$, the K41 prediction is 
\begin{equation} 
E^{K41}(k) \sim k^{-5/3} , 
\end{equation} 
a result that is in good agreement with a wide range of experiments [see,
e.g., Refs.~\cite{frisch,sreeniantonia}].

The structure functions $S_p(r)$ are the moments of the PDFs of the
longitudinal velocity increments $\delta u_{||} \equiv [(u({\bf x} + {\bf
r}) - u({\bf x}))\cdot({\bf r}/r)] $. [In the argument of $S_p$ we use $r$
instead of ${\bf r}$ when we consider homogeneous, isotropic turbulence.]
These PDFs have been measured directly~\cite{prasonc} and they show
non-Gaussian tails; as $r$ decreases, the deviations of these PDFs from
Gaussian distributions increases.

We now present a few examples of recent Lagrangian
measurements~\cite{pintonacc,bodenacc} that have been designed to track
tracer particles in, e.g., the von K\'arm\'an flow at large Reynolds
numbers.  By employing state-of-the-art measurement techniques, such as
silicon strip detectors~\cite{bodenacc}, used in high-energy-physics
experiments, or acoustic-doppler methods~\cite{pintonacc}, these experiments
have been able to attain high spatial resolution and high sampling rates
and have, therefore, been enable to obtain good data for acceleration
statistics of Lagrangian particles and the analogues of velocity structure
functions for them.  

These experiments~\cite{bodenacc} find, for $500 < Re_{\lambda} <
970$, consistency with the
Heisenberg-Yaglom scaling form of the acceleration variance, i.e., 
\begin{equation}
\langle a_i a_j \rangle 
\sim \epsilon^{(3/2)} \nu^{(-1/2)} \delta_{ij},
\end{equation}
where $a_i$ is the Cartesian component $i$ of the acceleration.
Furthermore, there are indications of strong intermittency effects
in the acceleration of particles and anisotropy effects are
present even at very large $Re_{\lambda}$. 

Order-$p$ Lagrangian velocity structure functions are defined along a
Lagrangian trajectory as \begin{equation} S^L_{i,p}(\tau) = \langle
[v^L_i(t + \tau) - v^L_i(t)]^p\rangle, \end{equation} where the
superscript $L$ denotes Lagrangian and the subscript $i$ the Cartesian
component. If the time lag $\tau$ lies in the temporal analogue of the
inertial range, i.e., $\tau_\eta \ll \tau \ll T_L$, where $\tau_\eta$ is
the  viscous dissipation time scale and $T_L$ is the time associated with
the  scale $L$ at which energy is injected into the system, then it is
expected that 
\begin{equation} 
S^L_{i,p}(\tau) \sim \tau^{\zeta^L_{i,p}}. 
\end{equation}
The analogue of the dimensional K41 prediction is $\zeta^{L,K41}_{i,p} =
p/2$; experiments and simulations~\cite{ictr1} indicate that there are
corrections to this simple dimensional prediction.

The best laboratory realisations of  two-dimensional turbulence are (a) a
thin layer of a conducting fluid excited by magnetic fields, varying both in
space and time and applied perpendicular to the layer~\cite{sommeria}, and
(b) soap films~\cite{couder} in which turbulence can be generated either by
electromagnetic forcing or by the introduction of a comb, which plays the
role of a grid, in a rapidly flowing soap film. In the range of parameters
used in typical experimental
studies~\cite{ecke,tabelingrev,goldburgrev,rivera} both these systems can be
described quite well~\cite{chomaz,prasad02} by the 2D Navier Stokes equation
(see Sec. 3) with an additional Ekman-friction term, induced typically by
air drag; however, in some cases we must also account for corrections
arising from fluctuations of the film thickness, compressibility effects,
and the Marangoni effect.  Measurement techniques are similar to those
employed to study three-dimensional
turbulence~\cite{ecke,tabelingrev,goldburgrev}. Two-dimensional analogues of
the PDFs described above for 3D turbulence have been measured [see, e.g.,
Refs.~\cite{rivera}]; we will touch on these briefly when we discuss
numerical simulations of 2D turbulence in Sec. 6.3. Velocity and vorticity
structure functions can be measured as in 3D turbulence; however, inertial
ranges associated with inverse and forward cascades must be distinguished;
the former shows simple scaling with an energy spectrum $E(k) \sim k^{-5/3}$
whereas the latter has an energy spectrum $E(k) \sim k^{-(3+\delta)}$, with
$\delta = 0$ if there is no Ekman friction and $\delta > 0$ otherwise.  In
the forward cascade velocity structure functions show simple
scaling~\cite{rivera}; we are not aware of experimental measurements of
vorticity structure functions (we will discuss these in the context of
numerical simulations in Sec. 6.3).

We end this Section with a brief discussion of one example of turbulence in
a non-Newtonian setting, namely, fluid flow in the presence of polymer
additives. There are two dimensionless control parameters in this case: $Re$
and the Weissenberg number $We$, which is a ratio of the polymer-relaxation
time and a typical shearing time in the flow (some
studies~\cite{procdragrev} use a similar dimensionless parameter called the
Deborah number $De$). Dramatically different behaviours arise depending on
the values of these parameters. 

In the absence of polymers the flow is  laminar at low $Re$; however, the
addition of small amounts of  high-molecular-weight polymers can induce {\it
elastic turbulence}~\cite{vsnjp}, i.e., a mixing flow that is like
turbulence and in which the drag increases with increasing $We$. We will not
discuss elastic turbulence in detail here; we refer the reader to
Ref.~\cite{vsnjp} for an overview of experiments and to
Ref.~\cite{2delastic} for representative numerical simulations. 

If, instead, the flow is turbulent in the absence of polymers, i.e., we
consider large-$Re$ flows, then the addition of polymers leads to the
dramatic phenomenon of {\it drag reduction} that has been known since
1949~\cite{toms}; it has obvious and important industrial applications
\cite{lumleyrev,procdragrev,hoyt,virk,krswhite}.  Normally
drag reduction is discussed in the context of pipe or channel flows: on the
addition of polymers to turbulent flow in a pipe, the pressure difference
required to maintain a given volumetric flow rate {\it decreases}, i.e., the
drag is reduced and a percentage drag reduction can be obtained from the
percentage reduction in the pressure difference. For a recent discussion of
drag reduction in pipe or channel flows we refer the reader to
Ref.~\cite{procdragrev}. Here we concentrate on other phenomena that are
associated with the addition of polymers to turbulent flows that are
homogeneous and isotropic. In particular, experiments~\cite{hoyt} show that
the polymers lead to a suppression of small-scale structures and important
modifications in the second-order structure function~\cite{bodenpoly}.  We
will return to an examination of such phenomena when we discuss direct
numerical simulations in Sec. 6.5.

\section{Models} 

Before we discuss advances in the statistical characterization of
turbulence, we provide a brief description to the models we consider. We
start with the basic equations of hydrodynamics, in three and two
dimensions, that are central to studies of turbulence. We also give
introductory overviews of the Burgers equation in one dimension, the
advection-diffusion equation for passive scalars, and the coupled NS and
finitely extensible nonlinear elastic Peterlin (FENE-P) equations for
polymers in a fluid.  We end this Section with a description of shell models
that are often used as highly simplified models for homogeneous, isotropic
turbulence.

At low Mach numbers, fluid flows are governed by the Navier-Stokes (NS)
Eq. (7) augmented by the incompressibility condition 
\begin{eqnarray} 
\partial_t{\bf u} +
({\bf u.}\nabla){\bf u} &=& -\nabla p + \nu\nabla^2{\bf u} + {\bf
f}, \nonumber \\
\nabla \cdot {\bf u} &=& 0, 
\label{NS} 
\end{eqnarray} 
where we use units in which the density $\rho$ = 1, the Eulerian velocity
at point ${\bf r}$ and time $t$ is ${\bf u}({\bf r},t)$, the external body
force per unit volume is ${\bf f}$ , and  $\nu$ is the kinematic viscosity. 
The pressure $p$ can be eliminated by using the incompressibility 
condition~\cite{pope,frisch,batchelor} and it can then be obtained from the
Poisson equation $\nabla^2 p = -\partial_{ij}(u_iu_j)$.  In the unforced,
inviscid case, the momentum, the kinetic energy, and the helicity $H \equiv
\int d {\bf r} \omega \cdot {\bf u}/2 $ are conserved; here $\omega \equiv
\nabla \times {\bf u}$ is the vorticity.  The Reynolds number $Re \equiv
LV/\nu$, where $L$ and $V$ are characteristic length and velocity scales, is
a convenient dimensionless control parameter:  The flow is laminar at low
$Re$ and irregular, and eventually turbulent, as $Re$ is increased.  
 
In the vorticity formulation the NS equation~\ref{NS} becomes 
\begin{equation} 
\partial_t{\bf \omega} = \nabla \times {\bf
u}\times {\bf \omega} + \nu\nabla^2{\bf \omega} + \nabla \times
{\bf f};  
\end{equation} 
the pressure is eliminated naturally here.  This formulation is particularly
useful is two dimensions since $\omega$ is a pseudo-scalar in this case. 
Specifically, in two dimensions, the NS equation can be written in terms of 
$\omega$ and the stream function $\psi$:  
\begin{eqnarray}
\partial_t \omega - J(\psi,\omega) &=& \nu \nabla^2 \omega + \alpha_E\omega + f;\nonumber \\ 
\nabla^2 \psi &=& \omega; \nonumber \\ 
J(\psi,\omega) &\equiv& (\partial_x
\psi)(\partial_y \omega)
                  - (\partial_x \omega) (\partial_y \psi).
\end{eqnarray}
Here $\alpha_E$ is the coefficient of the air-drag-induced 
Ekman-friction term. The incompressibility constraint
\begin{equation}
 \partial_x u_x + \partial_y u_y = 0
\end{equation} 
ensures that the velocity is uniquely determined by $\psi$  via
\begin{equation} 
{\bf u } \equiv (-\partial_y \psi, \partial_x \psi).  
\end{equation}
In the inviscid, unforced case we have more conserved quantities in  two
dimensions than in three; the additional conserved quantities are $\langle
\frac{1}{2}\omega^n \rangle$, for all powers $n$, the first of which is the 
mean enstrophy, $\Omega = \langle \frac{1}{2}\omega^2 \rangle$. 

In one dimension (1D) the incompressibility constraint leads to trivial
velocity fields. It is fruitful, however, to consider the Burgers
equation~\cite{burgers}, which is the NS equation without pressure and the
incompressibility constraint.  This has been studied in great detail as it
often provides interesting insights into fluid turbulence. In 1D the Burgers
equation is 
\begin{equation} 
\partial_t v +v\partial_x v =\nu
\nabla^2 v + f, 
\label{burgers} 
\end{equation} 
where $f$ is the external force and the velocity $v$ can have shocks since 
the system is compressible.  In the unforced, inviscid case the Burgers 
equation has infinitely many conserved quantities, namely, $\int v^n dx$ for
all integers $n$. In the limit $\nu \rightarrow 0$ we can use the Cole-Hopf
transformation, $v = \partial_x \Psi$, $f \equiv -\partial_x F$, and $\Psi
\equiv 2\nu \ln \Theta$, to obtain $\partial_t \Theta=\nu\partial_x^2\Theta +
F\Theta/(2\nu)$, a linear partial differential equation (PDE) that can be
solved explicitly in the absence of any boundary~\cite{fri00,bec07}.

Passive scalars such as pollutants can be advected by fluids. These flows
are governed by the advection-diffusion equation 
\begin{equation} 
\partial_t \theta + {\bf u}.\nabla
\theta = \kappa\nabla^2\theta + {\bf f}_\theta, 
\label{adv-diff}
\end{equation} 
where $\theta$ is the passive-scalar field, the advecting velocity field
{\bf u} satisfies  the NS equation~\ref{NS}, and ${\bf f}_\theta$ is an
external force.  The field $\theta$ is {\it passive} because it does not act
on or modify ${\bf u}$. Note that Eq.(~\ref{adv-diff}) is linear in
$\theta$. It is possible, therefore, to make considerable analytical
progress in understanding the statistical properties of passive-scalar
turbulence for the simplified model of passive-scalar advection due to
Kraichnan~\cite{falcormp,kraichpass}; in this model each component of ${\bf
f}_\theta$ is a zero-mean Gaussian random variable that is white in time;
furthermore, each component of ${\bf u}$ is taken to be a zero-mean Gaussian
random variable that is white in time and which has the covariance
\begin{equation}
\la u_i({\bf x},t)u_j({\bf x} + {\bf r},t^{\prime})\ra = 2D_{ij}\delta(t-t^
{\prime});
\label{covarnce}
\end{equation}
the Fourier transform of $D_{ij}$ has the form
\begin{equation}
\tilde D_{ij}({\bf q}) \propto {\big (}q^2 + \frac{1}{L^2}{\big )}^{-(d +
\xi)/2}e^{-\eta q^2}
{\big [}\delta_{ij} - \frac{q_iq_j}{q^2}{\big ]};
\end{equation}
{\bf q} is the wave vector, $L$ is the characteristic large length scale, $
\eta$ is the dissipation scale, and $\xi$ is a parameter. In the limit of $L
\to \infty$ and $\eta \to 0$ we have, in real space,
\begin{equation}
D_{ij}({\bf r}) = D^0\delta_{ij} - \frac {1}{2}d_{ij}({\bf r})
\label{}
\end{equation}
with
\begin{equation}
d_{ij} = D_1r^\xi\big [(d-1+\xi)\delta_{ij} -\xi\frac{r_ir_j}{r^2}\big ].
\end{equation}
$D_1$ is a normalization constant and $\xi$ a parameter; for $0 < \xi < 2$ 
equal-time passive-scalar structure functions show 
multiscaling~\cite{falcormp}.

We turn now to an example of a model for non-Newtonian flows. This model
combines the NS equation for a fluid with the finitely extensible nonlinear
elastic Peterlin (FENE-P) model for polymers; it is used {\it inter alia} to
study the effects of polymer additives on fluid turbulence. This model is
defined by the following equations: 
\begin{eqnarray} 
\partial_t{\bf u} + ({\bf u}.{\bf
\nabla}){\bf u} &=& \nu \nabla^2 {\bf u}+
              \frac{\mu}{\tau_P}\nabla.[f(r_P){\cal C}]
              - {\nabla}p;
                                                 \label{ns}\\
\partial_t {\cal C} + {\bf u}.{\bf \nabla}{\cal C} &=& {\cal
C}. (\nabla {\bf u}) +
                {(\nabla {\bf u})^T}.{\cal C} -
                \frac{{f(r_P){\cal C} }- {\cal I}}{\tau_P}.
\label{FENE}
\end{eqnarray} 
Here $\nu$ is the kinematic viscosity of the fluid, $\mu$ the viscosity
parameter for the solute (FENE-P), $\tau_P$ the polymer relaxation time,
$\rho$ the solvent density, $p$ the pressure, $(\nabla {\bf u})^T$ the
transpose of  $({\nabla {\bf u}})$, ${\cal C}_{\alpha\beta}\equiv
{\langle{R_\alpha}{R_\beta}\rangle}$ the elements of the
polymer-conformation tensor ${\cal C}$ (angular brackets indicate an average
over polymer configurations), ${\cal I}$ the identity tensor with elements
$\delta_{\alpha \beta}$, $f(r_P)\equiv{(L^2 -3)/(L^2 - r_P^2)}$ the FENE-P
potential that ensures finite extensibility, $ r_P \equiv \sqrt{Tr(\cal C)}$
and $ L $ the length and the maximum possible extension, respectively, of
the polymers, and $c\equiv\mu/(\nu+\mu)$ a dimensionless measure of the
polymer concentration~\cite{vai03}.

The hydrodynamical partial differential equations (PDEs) discussed above are
difficult to solve, even on computers via direct numerical simulation (DNS),
if we want to resolve the large ranges of spatial and temporal scales that
become relevant in turbulent flows. It is useful, therefore, to consider
simplified models of turbulence that are numerically more tractable than
these PDEs.  {\it Shell models} are important examples of such simplified
models; they have proved to be useful testing grounds for the multiscaling
properties of structure functions in turbulence. We will  consider, as
illustrative examples, the Gledzer-Ohkitani-Yamada (GOY) shell
model~\cite{goy} for fluid turbulence in three dimensions and a  shell model
for the advection-diffusion equation~\cite{adshell}.

Shell models cannot be derived from the NS equation in any systematic way.
They are formulated in a discretised Fourier space with logarithmically
spaced wave vectors $k_n= k_0 {\tilde \lambda}^n\/, {\tilde \lambda} > 1,$ associated with 
shells $n$ and dynamical variables that are the complex, scalar velocities
$u_n\/$. Note that $k_n$ is chosen to be a scalar: spherical symmetry is
implicit in GOY-type shell models since their aim is to study homogeneous,
isotropic turbulence. Given that $k_n$ and $u_n$ are scalars, shell models
cannot describe vortical structures or enforce the incompressibility 
constraint. 

The temporal evolution of such a shell model is governed by a set of
ordinary differential equations
that have the following features in common with the Fourier-space version of
the NS equation~\cite{lesieur}: they have a viscous-dissipation term  of
the form $-\nu k_n^2u_n$,  they conserve the shell-model analogues of the
energy and the helicity in the absence of viscosity and forcing,  and they
have nonlinear terms of the form $\imath k_nu_nu_{n'}$ that couple
velocities in different shells.  In the NS equation all Fourier modes of
the velocity affect each other directly but in most shell models nonlinear
terms limit direct interactions to  shell velocities in nearest-  and
next-nearest-neighbour shells; thus direct {\it sweeping effects}, i.e., the
advection of the largest eddies by the the smallest eddies, are present in
the NS equation but not in most shell models. This is why the latter are
occasionally viewed as a  highly simplified, quasi-Lagrangian representation
(see below) of the NS equation.

The GOY-model evolution equations have the form
\begin{eqnarray}
[ \frac{d}{dt} + \nu k_n^2]u_n =  i(a_n u_{n+1}u_{n+2} b_n u_{n-1}u_{n+1} + c_n
u_{n-1}u_{n-2} )^{\ast} + f_n, 
\label{goy}
\end{eqnarray}
where complex conjugation is denoted by $\ast$, the coefficients are chosen
to be $a_n = k_n$, $b_n = -\delta k_{n-1}$, 
$c_n = -(1-\delta)k_{n-2}$ to conserve the shell-model analogues of the
energy and the helicity in the inviscid, unforced case; in any practical 
calculation $1 \leq n \leq N$, where $N$ is the total number of shells and 
we use the boundary conditions $u_n = 0 \, \forall \, n < 1$ or 
$\forall \, n > N$; as
mentioned above $k_n = {\tilde \lambda}^n k_0$ and many groups use 
 ${\tilde \lambda} = 2\/$, $\delta = 1/2$, $k_0= 1/16$, and $N=22$. The logarithmic 
discretisation here allows 
us to reach very high Reynolds number, in numerical simulations of this 
model, even with such  a moderate value of  $N$. For studies of decaying
turbulence we set $f_n = 0, \forall \, n$; in the case of statistically
steady, forced turbulence~\cite{epjb} it is convenient to use 
$f_n = (1 + \imath) 5 \times 10^{-3}$. For such a shell model the 
analogue of a velocity structure function is 
$S_p(k_n) = \langle |u(k_n)|^p \rangle$ and the energy spectrum is 
$E(k_n) = |u(k_n)|^2/k_n$.

It is possible to construct other shell models, by using arguments 
similar to the ones we have just discussed, for other PDEs such as 
the advection-diffusion equation. Its shell-model version  is 
\begin {eqnarray}
[\frac{d}{dt} + \kappa k_n^2]\theta  = i[k_n(\theta_{n+1}u_{n-1} -
\theta_{n-1}u_{n+1}) - \nonumber\\\frac{k_{n-1}}{2}
(\theta_{n-1}u_{n-2} + \theta_{n-2}u_{n-1})
-\nonumber\\\frac{k_{n-1}}{2}(\theta_{n+2}u_{n+1} + \theta_{n+1}u_{n+2})]^*
\label{}
\end{eqnarray}

For this model, the advecting velocity field can either be obtained from the
numerical solution of a fluid shell model, like the GOY model above, or by
using a shell-model version of the type of stochastic velocity field
introduced in the Kraichnan model for passive-scalar advection~\cite{ssrnjp}.
A shell-model analogue for the FENE-P model of fluid turbulence with polymer
additives may be found in Ref.~\cite{chirag0}. 

\subsection{Eulerian, Lagrangian, Quasi-Lagrangian frameworks} 

The Navier-Stokes Eq.(~\ref{NS}) is written in terms of the Eulerian
velocity ${\bf u}$ at position ${\bf x}$ and time $t$; i.e., in the Eulerian
case  we use a frame of reference that is fixed with respect to the fluid;
this  frame  can be used for any flow property or field. The Lagrangian
framework~\cite{pope} uses a complementary point of view in which we fix a
frame of reference to a fluid  {\it particle}; this fictitious particle
moves with the flow and its path is known as a Lagrangian trajectory. Each
Lagrangian particle is characterised by its position vector ${\bf r_0}$ at
time $t_0$; its trajectory at some later time $t$ is ${\bf R} = {\bf
R}(t;{\bf r_0},t_0)$ and the associated Lagrangian velocity is 
\begin{equation} 
{\bf v} = \biggl (\frac{d{\bf R}}{dt}
\biggl )_{{\bf r_0}}.  
\end{equation}
We will also employ the quasi-Lagrangian \cite{belinicher,lvov} framework
that uses the following transformation for an Eulerian field $\psi({\bf
r},t)$: 
\begin{equation}
{\hat \psi}({\bf r},t) 
\equiv \psi[{\bf r} + {\bf R}(t;{\bf r_0},0),t] ;
\label{qltrans}
\end{equation}
here ${\hat \psi}$ is the quasi-Lagrangian field and ${\bf R}(t;{\bf
r_0},0)$ is the position at time $t$ of a  Lagrangian particle that was at
point ${\bf r_0}$ at time $t = 0$.

As we have mentioned above, sweeping effects are present when we use
Eulerian velocities. However, since Lagrangian particles move with the 
flow, such effects are not present in Lagrangian and quasi-Lagrangian
frameworks. In  experiments neutrally buoyant tracer particles are used 
to obtain Lagrangian trajectories that can be used to obtain statistical 
properties of Lagrangian particles.

\section{Homogeneous Isotropic Turbulence: Phenomenology}

In 1941 Kolmogorov~\cite{K41} developed his classic phenomenological
approach to turbulence that is often referred to as K41. He used the idea of
the Richardson cascade to provide an intuitive, though not rigorous,
understanding of the power-law behaviours we have mentioned in
Sec. 2.  We give a brief introduction to K41
phenomenology and related ideas; for a detailed discussion the reader should
consult Ref.~\cite{frisch}.

First we must recognise that there are two important length scales: (a) The
large {\it integral length scale} $L$ that is comparable to the system size
and at which energy injection takes place; flow at this scale depends on the
details of the system and the way in which energy is injected into it; (b)
and  the  small {\it dissipation length scale} $\eta_d$ below which energy
dissipation becomes significant. The  inertial range of scales, in which
structure functions and energy spectra assume the power-law behaviours
mentioned above (Sec. 2), lie in between $L$ and $\eta$; as $Re$ increases
so does the extent of the inertial range. 

In K41 Kolmogorov made the following assumptions: (a) Fully developed 3D
turbulence is homogeneous and isotropic at small length scales and far away
from boundaries. (b) In the statistical steady state, the energy dissipation
rate per unit volume $\epsilon$ remains finite and positive even when
$Re\to\infty$ (the dissipative anomaly mentioned above). (c) A
Richardson-type cascade is set up in which energy is transferred by the
breakdown of the largest eddies, created by inherent instabilities of the
flow, to smaller ones, which decay in turn into even smaller eddies, and so
on till the sizes of the eddies become comparable to $\eta_d$ where their
energy can then be degraded by viscous dissipation.  As $Re\to\infty$ all
inertial-range statistical properties are uniquely and universally
determined by the scale $r$ and $\epsilon$ and are independent of $L$, $\nu$
and $\eta_d$.  

Dimensional analysis then yields the scaling form of the
order-$p$ structure function 
\begin{equation}
S_p^{K41}(r) \approx  C\epsilon^{p/3}r^{p/3},
\label{K41eq1}
\end{equation}
since $\epsilon$ has dimensions of $(length)^2(time)^{-3}$.
[It is implicit here that the eddies, at any given level of the Richardson
cascade, are space filling; if not, $\epsilon$ is intermittent 
and scale dependent as we discuss in Sec. 5 on multiscaling.] 
Thus $\zeta_p^{K41} = p/3$; for $p = 2$ we get $S_2^{K41}(r) \sim r^{2/3}$
whose Fourier transform is related to the K41 energy spectrum 
$E(k)^{K41}\sim k^{-5/3}$ (left panel of Fig. 1). 

The prediction $\zeta_3^{K41} = 1$, unlike all others K41 results, can
be derived exactly for the NS equation in the limit $Re\to\infty$. In
particular, it can be shown that~\cite{frisch,moninyaglom} 
\begin{equation}
S_3(\ell) \approx  -\frac{4}{5}\epsilon\ell,
\label{K41eq2}
\end{equation}
an important result, since it is both exact and nontrivial. 

It is often useful to discuss K41 phenomenology by introducing
$v_{\ell}$, the velocity associated with the inertial-range 
length scale $\ell$; clearly
\begin{equation}
v_{\ell} \sim \epsilon^{1/3}\ell^{1/3}.
\label{}
\end{equation}
The time scale $t_{\ell} \sim \frac{\ell}{v_{\ell}}$, the typical time
required for the transfer of energy from scales of order $\ell$ to smaller 
ones. This yields the rate of energy transfer 
\begin{equation}
\Pi \sim \frac{v^2_{\ell}}{t_{\ell}} \sim \frac{v^3_{\ell}}{\ell}.
\label{}
\end{equation}  
Given the assumptions of K41, there is neither direct energy injection nor
molecular dissipation in the inertial range. Therefore, the energy flux
$\Pi$ becomes independent of $\ell$ and is equal to the mean energy
dissipation rate $\epsilon$, which can now be written as
\begin{equation}
\epsilon \sim v^3_{\ell}/\ell. 
\label{vlk41}
\end{equation}

A similar prediction, for the two-point correlations of a passive-scalar
advected by a turbulent fluid is due to Obukhov and Corsin; we shall not
discuss it here but refer the reader to Ref.~\cite{obu,corr}.

\begin{figure}[htbp]
\epsfxsize=6cm
\centerline{\epsfbox{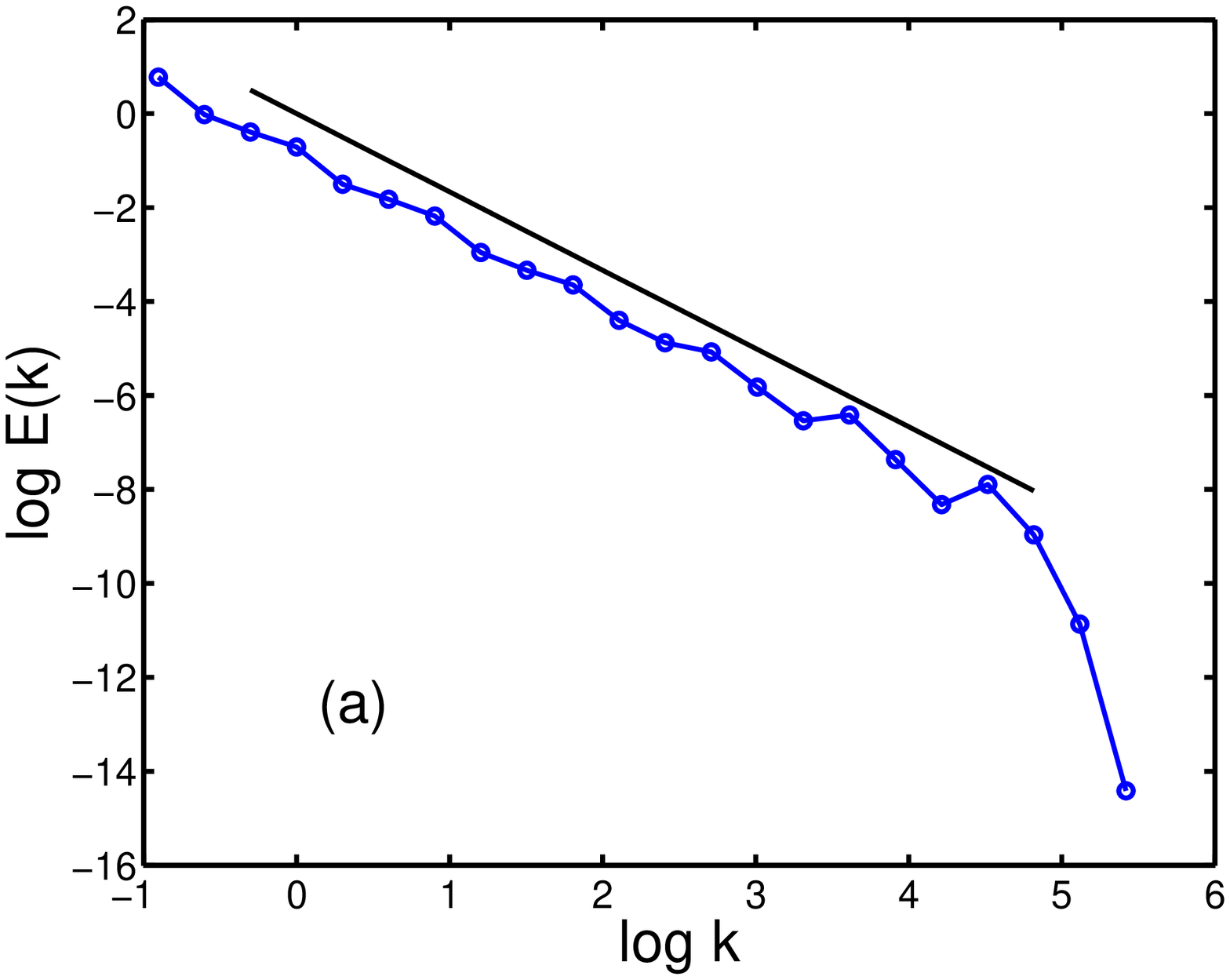},\epsfxsize=6cm
\epsfbox{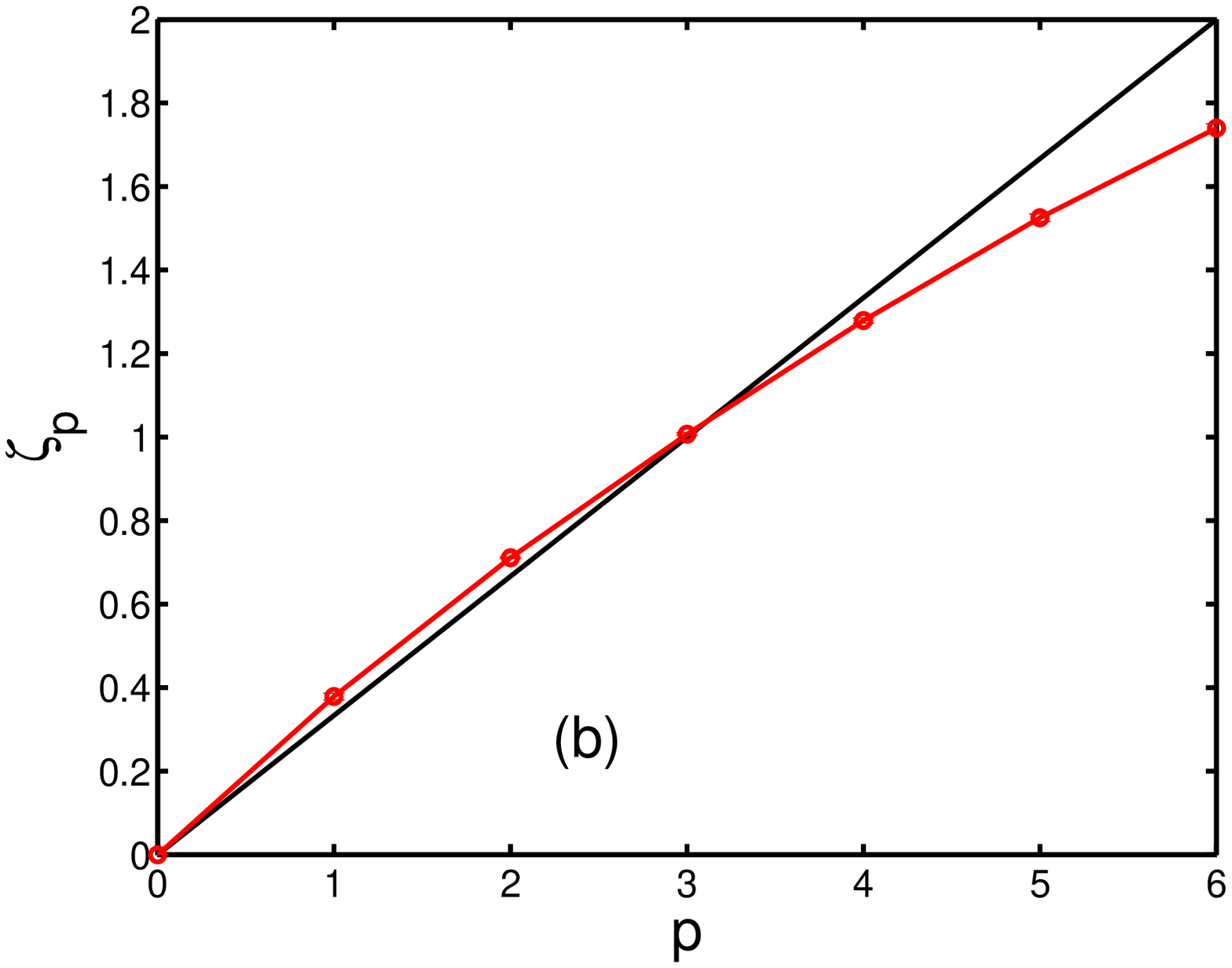}} 
\caption{(Color online) (a) A representative log-log plot of the 
energy spectrum $E(k)$ versus $k$, from 
a numerical simulation of the GOY shell model with 22 shells. The
straight black line is a guide to the eye indicating
K41 scaling $k^{-5/3}$. (b) A plot of the equal-time scaling
exponents $\zeta_p$ versus $p$, with error bars, obtained from the
GOY shell model. The straight black line (color online) indicates
K41 scaling $p/3$.}
\label{ekfig}
\end{figure}

As we have mentioned above, the cascade of energy in 3D turbulence
is replaced in 2D turbulence by a dual cascade: an inverse cascade
of energy from the injection scale to larger length scales and a
forward cascade of
enstrophy~\cite{tabelingrev,goldburgrev,kraichmont,kraic67}. In the
inverse cascade the energy accumulation at large length scales is
controlled eventually by Ekman friction. The analogue of K41
phenomenology for this case is based upon physical arguments due to
Kraichnan, Leith and Batchelor~\cite{kraic67}.  Given that there is
energy injection at some intermediate length scale, kinetic energy
get redistributed from such intermediate scales to the largest
length scale.  The scaling result for the two cascades gives us a
kinetic energy spectrum that has a $k^{-5/3}$ form in the
inverse-cascade inertial range and a $k^{-3}$ form (in the absence
of Ekman friction) in the forward-cascade inertial range.

\section{From scaling to multiscaling}

In equilibrium statistical mechanics, equal-time and time-dependent
correlation functions, in the vicinity of a critical point, display scaling
properties that are well understood. For example, for a spin system in $d$
dimensions close to its critical point, the scaling forms of the equal-time
correlation function $g(r;\bar{t},h)$ and its Fourier transform 
$\tilde g(k;\bar{t},h)$, for a pair of spins separated by a distance $r$, 
are as follows:
\begin{equation}
g(r;\bar{t},h) \approx  
\frac{G(r\bar{t}^{(\bar{\nu})},h/\bar{t}^{(\bar{\Delta})})}{r^{d-2+\bar{\eta}}};
\label{corr1}
\end{equation}
\begin{equation}
\tilde g(k;\bar{t},h) \approx 
\frac{\tilde G(k/\bar{t}^{(\bar{\nu})},h/\bar{t}^{(\bar{\Delta})})}{k^{2-
\bar{\eta}}}.
\label{corr2}
\end{equation}
Here the reduced temperature $\bar{t}= (T - T_c)/T_c$, where $T$ and $T_c$
are, respectively, the temperature and the critical temperature, and the
reduced field $h = H/k_BT_c$, with $H$ the external field and $k_B$ the
Boltzmann constant.  The equal-time critical exponents $\bar{\eta}$,
$\bar{\nu}$ and $ \bar{\Delta}$ are universal for a given universality class
(the unconventional overbars are used to distinguish these exponents from
the kinematic viscosity, etc.). The scaling functions $G$ and $\tilde{G}$
can be made universal too if two scale factors are taken into
account~\cite{wortis}. Precisely at the critical point $(\bar{t}=0, h=0)$
these scaling forms lead to power-law decays of correlation functions; and,
as the critical point is approached, the correlation length $\xi$ diverges
[e.g., as $\xi \sim \bar{t}^{(-\bar{\nu})}$ if $h = 0$]. Time-dependent
correlation functions also display scaling behaviour; e.g., the frequency
($\omega$) dependent correlation function has the scaling form to
Eq. (\ref{corr2}).
\begin{equation} 
{\tilde {\rm g}}(k,\omega;\bar{t},h) \approx \frac {\tilde
{\cal G}(k^{-z}\omega,k/\bar{t}^{(\bar{\nu})},h/\bar{t}^{(\bar{\Delta})})}{k^{2-\bar{\eta}}}.
\label{corr3}
\end{equation}
This scaling behaviour is associated with the divergence of
the relaxation time 
\begin{equation} 
\tau \sim \xi^z, 
\label{corr4}
\end{equation}
referred to as critical slowing down; here $z$ is the dynamic scaling
exponent.

In most critical phenomena in equilibrium statistical mechanics we obtain
the simple scaling forms summarised in the previous paragraph.  The
inertial-range behaviours of structure functions in turbulence (Secs. 2 and
3) are similar to the power-law forms of these critical-point correlation
functions. This similarity is especially strong at the level of K41 scaling
(Sec. 4); however, as we have mentioned earlier, experimental and numerical
work suggests significant {\it multiscaling} corrections to K41 scaling with
the equal-time multiscaling exponents $\zeta_p \neq \zeta_p^{K41}$; in
brief, multiscaling implies that $\zeta_p$ is not a linear function $p$;
indeed~\cite{frisch} it is a monotone increasing nonlinear function of $p$
(see right panel of Fig. 1).  The multiscaling of equal-time structure
functions seems to be a common property of various forms of turbulence,
e.g., 3D turbulence and passive-scalar turbulence. 

The multifractal model~\cite{frisch,parisifrisch,boffmazvulp} provides a way
of rationalising multiscaling corrections to K41. First we must give up
the K41 assumption of only one relevant length scale $\ell$ and 
the simple scaling form of Eq.(~\ref{vlk41}).
Thus we write the equal-time structure function as 
\begin{equation}
S_p(\ell) = C_p(\epsilon\ell)^{p/3}(\frac{\ell}{L})^{\delta_p},
\label{}
\end{equation}
where $\delta_p \equiv \zeta_p - p/3$ is the anomalous part of the scaling
exponent. 
We start with the assumption that the turbulent flow possesses a range of 
scaling exponents $h$ in the set $I = (h_{min},h_{max})$. For each $h$ in this
range, there 
is a set $\Sigma_h$ (in real space) of fractal dimension $D(h)$, such that, 
as $\ell/L \rightarrow 0$, 
\begin{equation}
\delta v_{\ell}({\bf r}) \sim \ell^h, 
\label{}
\end{equation} 
if ${\bf r} \in \Sigma_h$.
The exponents $(h_{min},h_{max})$ are postulated to be independent of the 
mechanism responsible for the turbulence. Hence  
\begin{equation}
S_p(\ell) \sim \int_I d\mu(h) (\ell/L)^{ph + 3 -D(h)},
\label{}
\end{equation}
where the $ph$ term comes from $p$ factors of $(\ell/L)$ in Eq. (34) and
the 
$3 - D(h)$ term comes from an additional factor of $(\ell/L)^{3-D(h)}$, which 
is the probability of being within a distance of $\sim \ell$ of the set
$\Sigma_h$ of 
dimension $D(h)$ that is embedded in three dimensions.  The co-dimension $D(h)$
and the 
exponents $h_{min}$ and $h_{max}$ are assumed to be universal 
\cite{frisch}. 
The measure $d\mu(h)$ gives the weight of the different
exponents. 
In the limit $\ell/L \rightarrow 0$ the method of steepest descent yields
\begin{equation}
\zeta_p = inf_h[ph + 3 - D(h)].
\label{}
\end{equation}
The K41 result follows from Eq. (36) if we allow for only one value 
of $h$, namely,  $h = 1/3$ and set $D(h) = 3$. For more details
we 
refer the reader to~\cite{frisch,parisifrisch,boffmazvulp};
the extension to time-dependent structure functions is given in 
Refs.~\cite{epjb,ssrnjp,dmrpprl}.

Exact results for multiscaling can be obtained for the Kraichnan 
model of passive-scalar turbulence. We outline the essential steps 
below; details may be found in Ref.~\cite{falcormp}.

The second-order correlation function is defined as 
\begin{equation}
C_2(\bf l,t) = \langle \theta(\bf x,t) \theta(\bf x + \bf l,t) \rangle.
\end{equation}
Here the angular brackets denote averaging over the statistics of 
the velocity and the force which are assumed to be independent of
one another ~\cite{falcormp}.
This equation of motion 
\begin{equation}
\partial_t C_2 ({\bf l},t) = \langle \partial_t \theta({\bf x},t) \theta({\bf x}
+ {\bf l},t) \rangle + \langle \theta({\bf x},t) \partial_t \theta({\bf x} +
{\bf l},t) \rangle
\end{equation}
is easy to solve by first by using the advection-diffusion equation and then
using 
Gaussian averages to obtain~\cite{falcormp}
\begin{equation}
\partial_t C_2(l) = D_1l^{1-d}\partial_l[(d-1)l^{d-1+\xi}C_2(l)] + 2\kappa
l^{1-d}\partial_l[l^{d-1}\partial_l C_2(l)] + \Phi(\frac{l}{L_1}),
\end{equation}
where $\Phi(\frac{l}{L_1})$ is the spatial correlation of the force~\cite{falcormp}  
(notice that we now work with just the scalar $l$ for the 
isotropic case). 
In the stationary state the time derivative vanishes on the left 
hand side. We impose the boundary 
conditions that, as $l \rightarrow \infty$, $C_2(l) = 0$,  
and $C_2(l)$ remains finite when $l \rightarrow 0$, whence 
\begin{equation}
C_2(l) = \frac{1}{(d - 1)D_1} \int_l^\infty \frac{r^{1-d}}{r^\xi +
l_d^\xi}dr\int_0^r\Phi (\frac{r}{L_1})y^{d-1}dy.
\end{equation}
In the limit $l_d << l << L_1$, the second-order structure function has the
following
scaling form,
\begin{equation}
S_2(l) \equiv  2[C_2(0) - C_2(l)] \approx  \frac{2}{(2-\xi)(d-1)D_1}\Phi (0)l^{2-\xi}, 
\end{equation}
i.e., equal-time exponents $\zeta_2^\theta = 2 -\xi$; 
this result follows from dimensional arguments as well. For 
order-$p$ correlation functions the equivalent of Eq. (38) 
can be written symbolically as ~\cite{falcormp}
\begin{equation}
\partial_t C_p = -M_pC_p + D_pC_p + F \otimes C_{p-2}
\end{equation}
where the operator $M_p$ is determined by the advection term, $D_p$ is the 
dissipative operator, and $F$ is the spatial correlator of the force. In the
limit 
of vanishing diffusivity, and in stationary state, the above equation reduces to
\begin{equation}
M_pC_p = F \otimes C_{p-2}.
\end{equation}
The associated homogeneous and  
inhomogeneous equations can be solved separately. By assuming scaling behaviour, we can extract the 
scaling exponent from simple dimensional analysis (superscript 
$dim$) to obtain
\begin{equation}
\zeta_p^{dim} = \frac{p}{2}(2-\xi).
\end{equation}

The solution $Z_p(\lambda {\bf r}_1,\lambda {\bf r}_2 ... \lambda 
{\bf r}_p)$ of the homogeneous part of Eq. (43) are called the 
zero-mode of the operator 
$M_p$. The zero-modes have the scaling property 
\begin{equation}
Z_p(\lambda {\bf r}_1,\lambda {\bf r}_2 ... \lambda {\bf r}_p) \sim
\lambda^{\zeta_p^{zero}}Z_p({\bf r}_1,{\bf r}_2 ... {\bf r}_p).
\end{equation}
Their scaling exponents $\zeta_p^{zero}$ cannot be determined from
dimensional 
arguments. The exponents $\zeta_p^{zero}$ are also called anomalous exponents.
And 
for a particular order-$p$ the actual scaling exponent is
\begin{equation}
\zeta_p = {\rm min} (\zeta_p^{zero},\zeta_p^{dim})
\end{equation}
This is how multiscaling arises in Kraichnan model of passive-scalar 
advection.
The principal difficulty lies in solving the problem with a particular boundary
condition. In 
recent times the following results have been obtained: Although the scaling 
exponents for the zero-modes has not been obtained exactly for any $p$, except
for 
$p = 2$ (in which case the anomalous exponent is actually subdominant),
perturbative 
methods have yielded the anomalous exponents. Also, it has been shown that the 
multiscaling disappears for $\xi > 2$ or $\xi < 0$ and that, although the
scaling 
exponents are universal, the amplitudes depend on the force correlator and hence
the
structure functions themselves are not universal. These results have been well
supported by numerical simulations.

Several studies of the multiscaling of equal-time structure 
functions have been carried out as outlined above. By contrast 
there are fewer studies of the multiscaling of time-dependent 
structure functions. We give an illustrative example for the 
Kraichnan model of passive-scalar advection.
For simplicity, we look at the Eulerian second-order time-dependent 
structure function which is defined, in Fourier space, as 
~\cite{ssrnjp,dmpass} 
\begin{equation}
{\tilde {\mathcal F}}^{\theta}({\bf k},t_0,t) = \langle {\tilde \theta}(-{\bf
k},t_0){\tilde \theta}({\bf k},t) \rangle.
\end{equation}
In order to arrive at a scaling form for $\tilde {{\mathcal F}}({\bf k},t_0,t)$,
we 
look at its equation of motion:
\begin{equation}
\frac{\partial \tilde {{\mathcal F}^{\theta}}({\bf k},t_0,t)}{\partial t} =
\langle{\tilde \theta}(-{\bf k},t_0)\frac{\partial{\tilde \theta}({\bf
k},t)}{\partial t}\rangle.
\label{eqnofmotn1}
\end{equation}
A spatial Fourier transform of the advection-diffusion equation 
(13) yields
\begin{equation}
\frac{\partial\tilde {\theta}(\bf k)}{\partial t} = i\int k_ju_j({\bf q})\tilde
{\theta}({\bf k} - {\bf q})d^dq -
\kappa k_jk_j\tilde {\theta}({\bf k}),
\label{advFT}
\end{equation}
so (\ref{eqnofmotn1}) maybe expanded as 
\begin{equation}
\frac{d\tilde {{\mathcal F}}^{\theta}({\bf k},t_0,t)}{dt} =
ik_j\int\langle\tilde {\theta}(-{\bf k},t_0)u_j({\bf q})\tilde {\theta}({\bf
k}-{\bf q},t)\rangle d^dq 
- \kappa k_jk_j\langle\tilde {\theta}(-{\bf k},t_0)\tilde {\theta}({\bf
k},t)\rangle.
\end{equation}
The above equation is solved with the help of Gaussian averaging. The first
term 
reduces to 
\begin{equation}
\langle\tilde {\theta}(-{\bf k},t_0)u_j({\bf q})\tilde {\theta}({\bf k}-{\bf
q},t)\rangle = \int_0^{\infty}\langle u_j(t)u_i(t^{\prime})\rangle
\langle\tilde {\theta}(-{\bf k},t_0)\frac{\delta}{\delta u_i(t^{\prime})}\tilde
{\theta}({\bf k}-{\bf q},t^{\prime})\rangle dt^{\prime}.
\end{equation}
Equations (\ref{covarnce}) and (\ref{advFT}) yield 
\begin{equation}
\frac{d\tilde {{\mathcal F}}({\bf k},t_0,t)}{dt} =
-2k_ik_j\int_0^{\infty}D_{ij}d^dq\tilde {{\mathcal F}}({\bf k},t_0,t).
\end{equation}
Since $2\int_0^{\infty}D_{ij}d^dq = D^0(L) \sim L^{\xi}$, the
equation of 
motion of the second-order structure function for the Eulerian field becomes  
\begin{equation}
\frac{\partial {\mathcal F}^{\theta}(r,t_0,t)}{\partial t} =
L^{\xi}\frac{\partial^2 {\mathcal F}^{\theta}(r,t_0,t)}{\partial r^2},
\label{E}
\end{equation}
whence ~\cite{ssrnjp}
\begin{equation}
\tilde {{\mathcal F}}({\bf k},t_0,t) = \phi(k,t_0)e^{-k^2L^{\xi}t}.
\end{equation} 
Thus it is clear that within the Eulerian framework we get a simple dynamic
scaling 
exponent $z$ = 2.
 
A similar analysis for the quasi-Lagrangian time-dependent structure 
function ~\cite{ssrnjp} gives 
\begin{equation}
\frac{\partial {\mathcal F}(r,t_0,t)}{\partial t} = (D^0\delta_{ij} -
D_{ij})\frac{\partial {\mathcal F}(r,t_0,t)}{\partial r_i\partial r_j} \sim
d_{ij}\frac{\partial {\mathcal F}(r,t_0,t)}{\partial r_i\partial r_j}. 
\label{QL}
\end{equation}

A Fourier transform of Eq. (\ref {QL}) yields   
$\tilde {{\mathcal F}}({\bf k},t_0,t) \propto \exp[-t/\tau]$, where 
$\tau = k^{\xi - 2}$, which implies a simple dynamic scaling exponent $z = 2 -
\xi$ 
in the quasi-Lagrangian framework. In Sec. 6.2 we discuss dynamic 
scaling and multiscaling in shell models.

\section{Numerical Simulations}
Numerical studies of the models described in Sec. 3 have 
contributed greatly to our understanding of turbulence. In this 
Section we give illustrative numerical studies of the 3D 
Navier-Stokes equation (Sec. 6.1), GOY and advection-diffusion 
shell models (Sec. 6.2), the 2D Navier-Stokes equation (Sec. 6.3), 
the 1D Burgers equation (Sec. 6.4) and the FENE-P model for 
polymer additives in a fluid (Sec. 6.5).

\subsection{3D Navier-Stokes Turbulence}
We concentrate on the statistical properties of homogeneous, 
isotropic turbulence, so we restrict ourselves 
periodic boundary conditions. Even with these simple 
boundary conditions,   
simulating these flows is a challenging task as a wide range of  
length scales has to be resolved. Therefore, 
state-of-the-art numerical simulations use pseudo-spectral 
methods that solve the Navier-Stokes equations via Fast Fourier 
transforms ~\cite{vin91,canuto} typically on supercomputers. 
For a discussion on the
implementation of the pseudo-spectral method we
refer the reader to Refs.~\cite{vin91,canuto}. We
outline this method below:
(a) Time marching is done by using either a  second-order, slaved
Adams-Bashforth or a Runge-Kutta scheme ~\cite{nr}.
(b) In Fourier space the contribution of the viscous term is 
-$\nu k^2 {\bf u}$.
(c) To avoid the computational costs of evaluating the 
convolution because of the non-linear term, it is first
calculated in real space and then Fourier transformed; hence 
the name pseudo-spectral method.
(d) In Fourier space the discretized Navier-Stokes time evolution us
     \begin{eqnarray}
     \nonumber
     \hspace{-0.5cm}u^{n+1}=\exp(-\nu k^2 \delta t)u^{n}+
     \frac{1-\exp(-\nu k^2 \delta{t})}{\nu 
k^2}P_{ij}[(3/2){\mathcal N}^n-(1/2){\mathcal N}^{n-1}]
     \end{eqnarray}
     where $n$ is the  iteration number, ${\mathcal N}$ indicates 
the non-linear term, and $P_{ij}=(\delta_{ij}-k_ik_j/k^2)$ is the transverse
projector which guarantees incompressibility.
(e) To suppress aliasing errors we use a $2/3$ dealiasing scheme 
~\cite{canuto}.

We give illustrative results from a direct numerical simulation DNS 
with $1024^3$ that we have carried out. This study uses the 
stochastic forcing of ~\cite{espope} and has attained a Taylor 
microscale Reynolds number $Re_\lambda \sim 100$, where 
$Re_{\lambda}=u_{rms}\lambda/\nu$;  
$u_{rms}=\sqrt{2E/3}$ is the root-mean-square velocity 
and the Taylor microscale $\lambda=\sqrt{\sum E(k)/\sum k^2 E(k)}$.
For state-of-the-art simulations with up to $4096^3$ collocation 
points we refer the reader to Ref.\cite{kanedarev}. As we had 
mentioned in Sec. 2, regions of high vorticity are organised into 
slender tubes. These can be visualised by looking at isosurfaces 
of $|\omega|$ as shown in the representative plots of Figs 2 and 3. 
The right panel of Fig. 2 shows the PDF of $|\omega|$; this has a 
distinctly non-Gaussian tail. The structure of high-$|\omega|$ 
vorticity tubes shows up especially clearly in the plots of Fig. 3, 
the second and third panels of which show successively magnified 
images of the central part of the first panel (for a $4096^3$ 
version see Ref. ~\cite{kanedarev}). 
\begin{figure}[htbp]
\epsfxsize=6cm
\centerline{\epsfbox{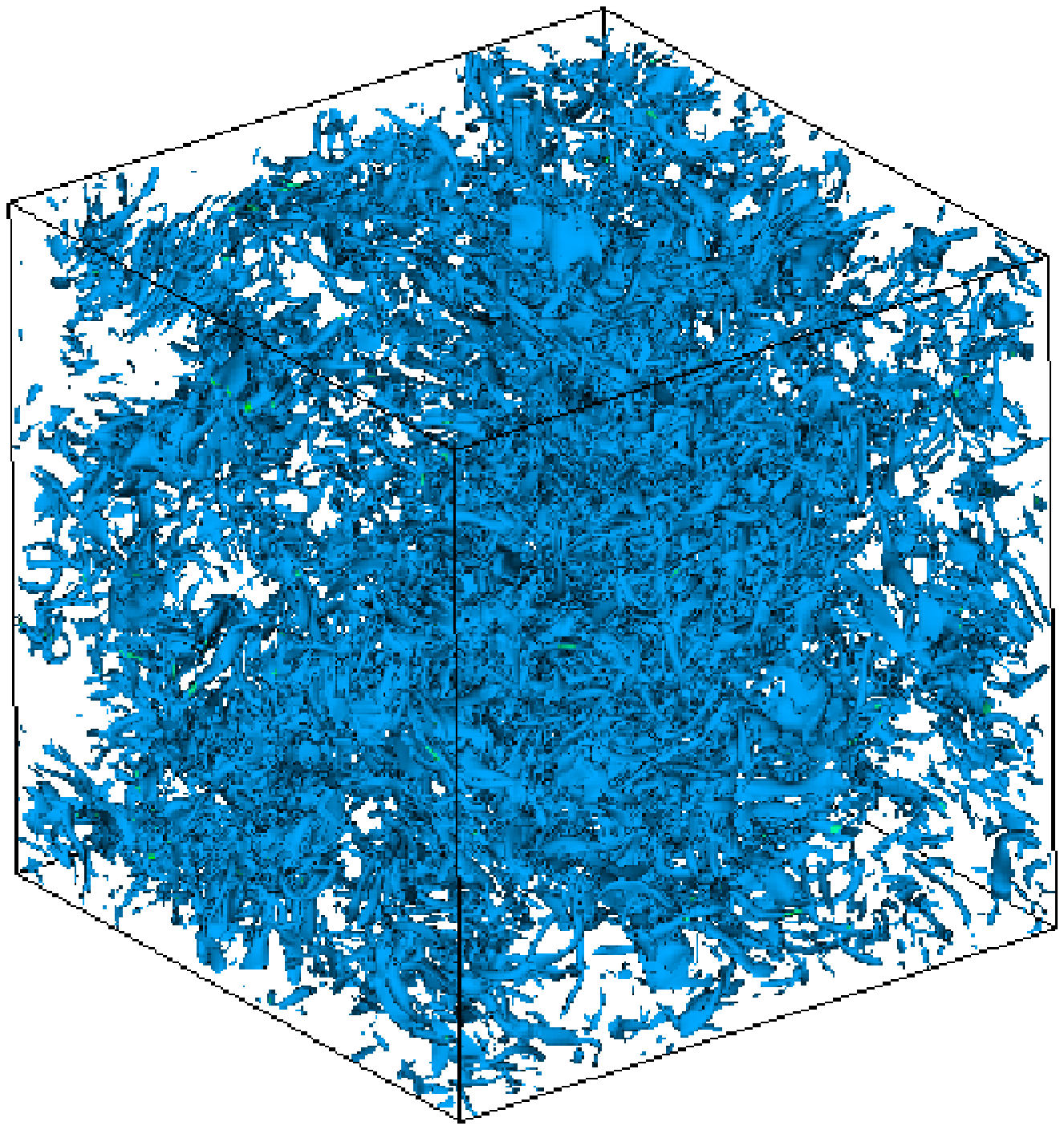}
\epsfxsize=6cm
\epsfbox{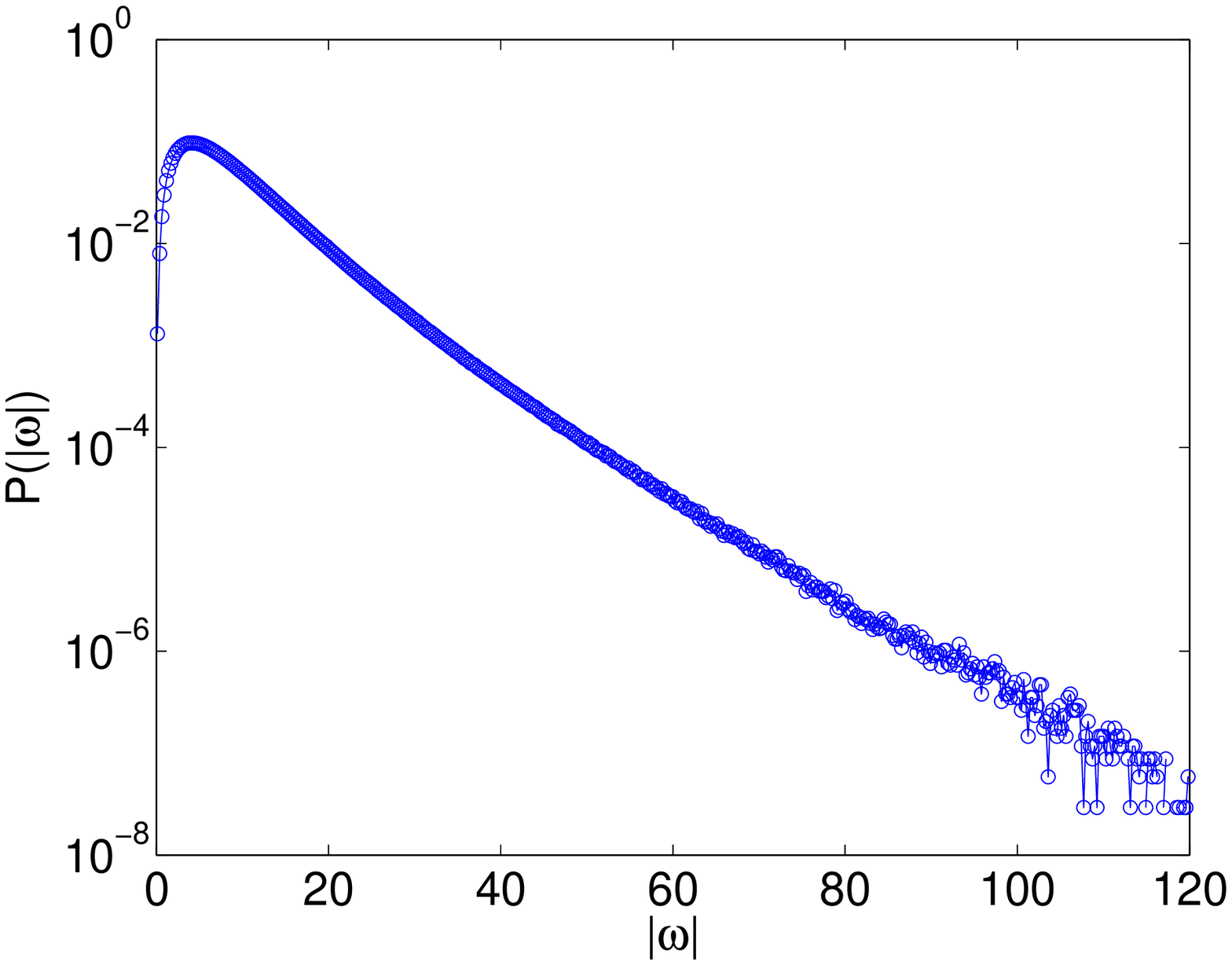}} 
\caption{(Color online) (Left) Isosurface plot of $|\omega|$ with 
$|\omega|$ equal to its mean value. (Right) A semilog plot of the 
PDF of $|\omega|$.}
\label{3diso1}
\end{figure}

\begin{figure}[htbp]
\epsfxsize=3.8cm
\centerline{\epsfbox{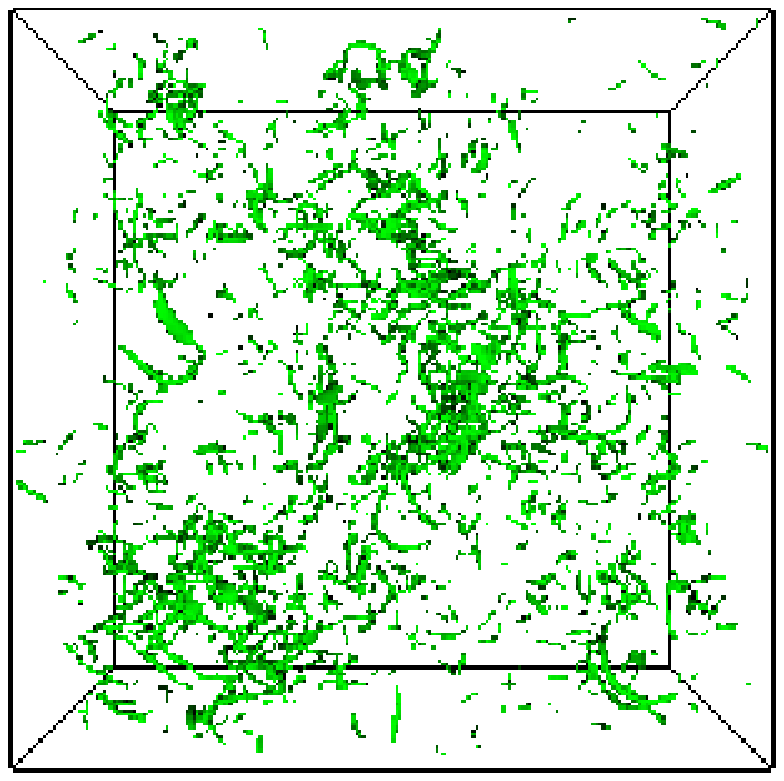}
\epsfxsize=3.8cm
\epsfbox{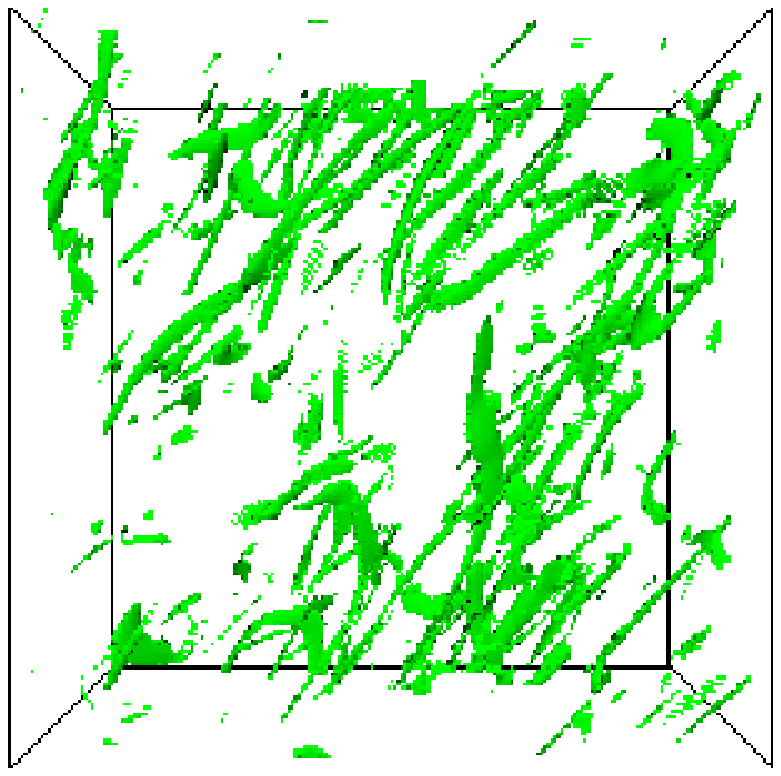} 
\epsfxsize=3.8cm
\epsfbox{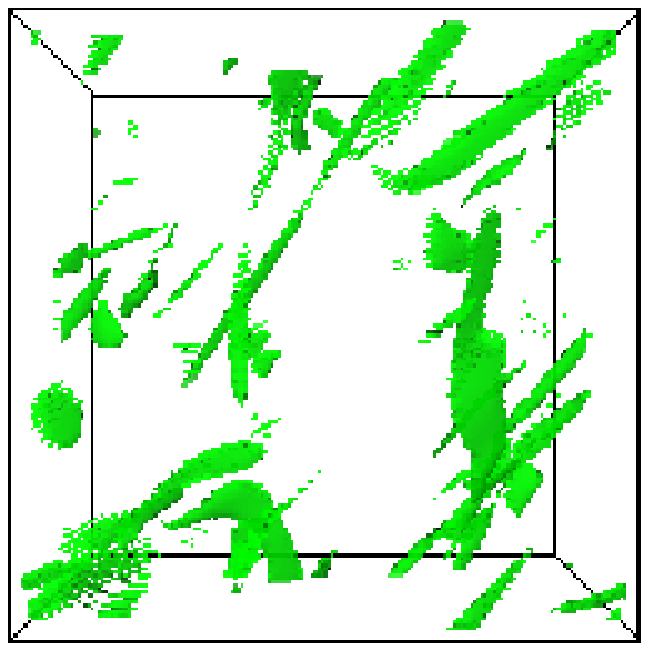}} 
\vspace{1cm}
\caption{(Color online) (Left) Isosurface plot of $|\omega|$ with 
$|\omega|$ equal to one standard deviation more than its mean 
value. (Center) A magnified version of the central part of the panel on the 
left.
(Right) A magnified version of the central part of the panel in the 
middle.}
\label{3diso2}
\end{figure}

\begin{figure}[htbp]
\epsfxsize=6cm
\centerline{\epsfbox{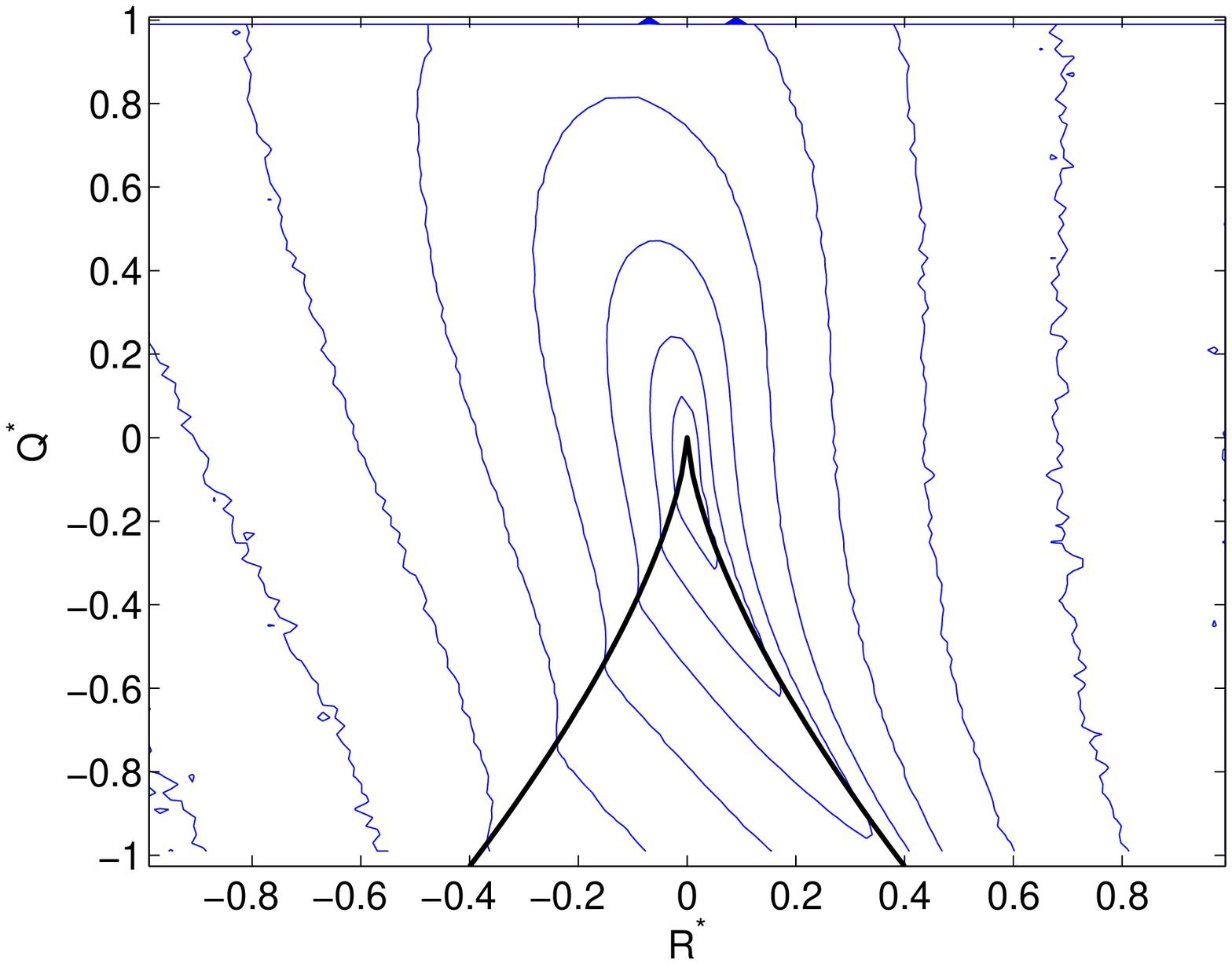} \epsfxsize=6cm
\epsfbox{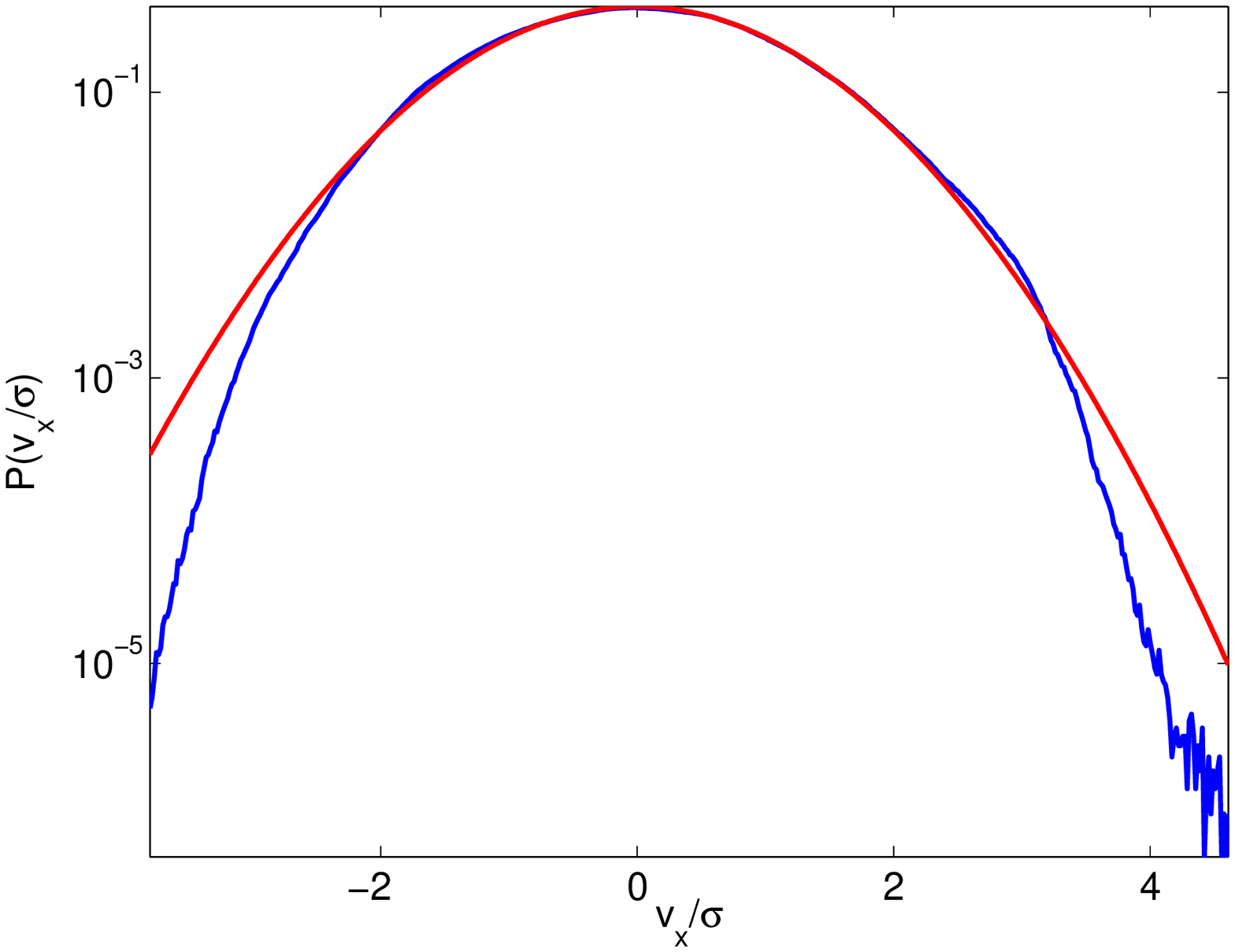}} 
\caption{(Color online) (Left) Joint PDF $P(Q^*,R^*)$ of $R^*=R/\langle s_{ij} s_{ij} 
\rangle^{3/2}$ and
$Q^*=Q/\langle s_{ij} s_{ij} \rangle$ calculated from our DNS. The black
curve represents the zero-discriminant (or Vieillefosse) line $27R^2/4+Q^3=0$.
The contour levels are logarithmically spaced.
(Right) PDF of the $x$-component of the velocity (here $\sigma$ 
denotes the standard deviation); the parabolic curve is a Gaussian 
that is drawn for comparison.}
\label{3diso1}
\end{figure}

\begin{figure}[htbp]
\epsfxsize=6cm
\centerline{\epsfbox{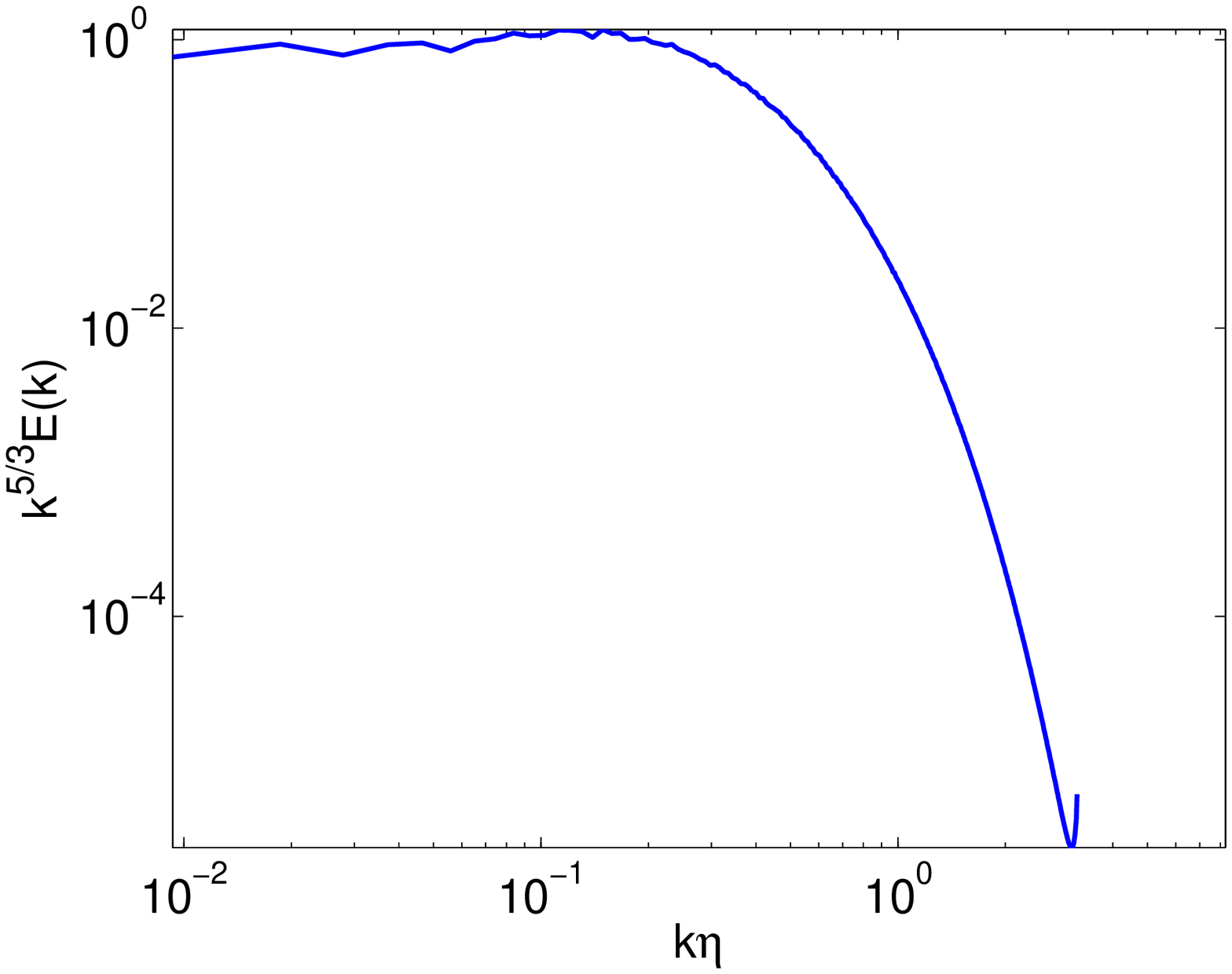} \epsfxsize=6cm \epsfbox{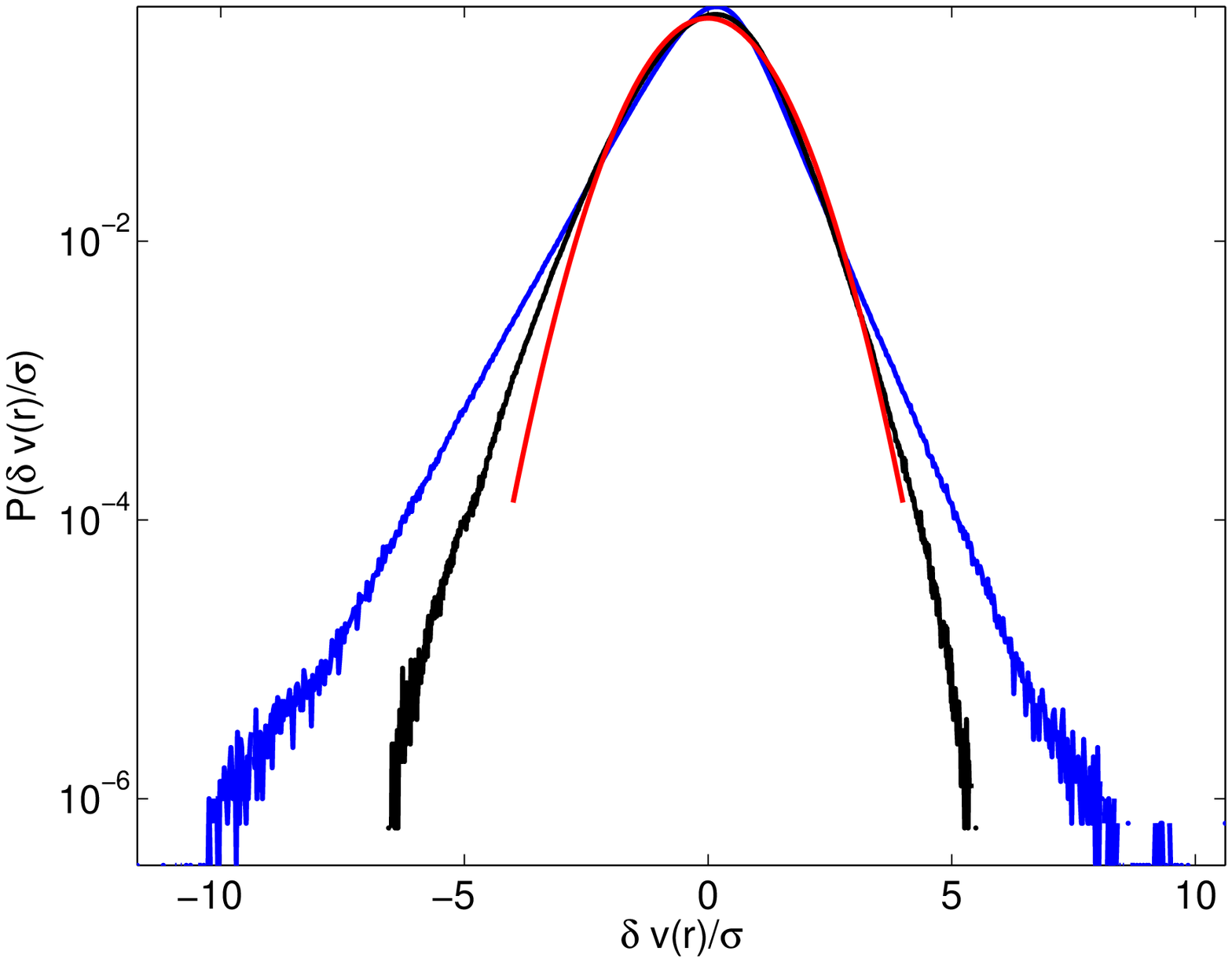}} 
\caption{(Color online) (Left) The compensated energy spectrum 
$k^{5/3}E(k)$ versus $k\eta$, where $\eta$ is the dissipation scale 
from our DNS (see text). (Right) PDFs of velocity 
increments that show marked deviation from Gaussian behaviour 
(innermost curve), especially at small length scales; the outermost 
PDF is for the velocity increment with the shorter length scale.}
\label{3diso1}
\end{figure}

One method to look at these structures is to study
the joint PDF of the invariants $Q=-tr(A^2)/2$ and $R=-tr(A^3)/3$
of the velocity gradient tensor.
The zero-discriminant or Vieillefosse line $D\equiv 27R^2/4 + Q^3 =0$ divides
the QR plane in different regions. The region with $D>0$ is vorticity dominant
(one of the eigenvalues of $A$ is greater than zero whereas the other
two eigenvalues are imaginary); the region $D<0$ is strain
dominated (all the eigenvalues of $A$ are real). The regions $D>0$ 
and $D<0$
can be further divided into two more quadrants depending upon
the sign of the eigenvalues. In the left panel of Fig. 4 we show a representative contour
plot of the joint PDF $P(Q^*,R^*)$ obtained from our DNS. The shape
of the contour is like a tear-drop, as in experiments 
~\cite{tsinober}, with a tail along the line $D=0$
in the region where $R^*>0$ and $Q^*<0$. The plot indicates that, 
in a numerical simulation, most of the structures are vortical but
there also exist regions of large strain.
For a more detailed discussion of the above
classification of different structures we refer the reader to
\cite{tsinober,cantwell}.

The left panel of Fig. 5 shows a plot of the compensated energy 
spectrum $k^{5/3}E(k)$ versus $k\eta$ ($\eta$ is the dissipation 
scale in our DNS). The flat portion at low $k\eta$ indicates 
agreement with the K41 form $E^{K41}(k) \sim k^{-5/3}$. There is 
a slight bump after that; this is referred to as a bottleneck 
(see Ref. ~\cite{kan03} and Sec 6.4); the spectrum then falls in 
the dissipation range. The right panel of Fig. 5 shows PDFs of 
velocity increments at different scales $r$. The innermost curve is 
a Gaussian for comparison; the non-Gaussian deviations increase as 
$r$ decreases.

We do not provide data for the multiscaling of velocity structure 
functions
in the 3D Navier-Stokes equation. We refer the reader to Ref.~\cite{ictr1}
for a recent discussion of such multiscaling. Often the inertial range is
quite limited in such studies. This range can be extended somewhat by using
the extended-self-similarity (ESS) procedure~\cite{ess} in which the slope of
a log-log plots of the structure function $S_{p}$ versus $S_{q}$ yields
the exponent ratio $\zeta_{p}/\zeta_{q}$; this procedure is especially
useful if $q=3$ since $\zeta_3 = 1$ for the 3D Navier-Stokes case.
We illustrate the use of this ESS procedure in Sec. (6.3) on 2D turbulence.

The methods of statistical field theory have been used with some success to
study the statistical properties of a randomly forced Navier-Stokes
equation~\cite{dommartin,yakors,adzhem,jkbbook}. The
stochastic force here acts at all length scales; it is Gaussian and
has a Fourier-space covariance proportional to $k^{1-y}$. For  $y \geq 0$,
a simple perturbation theory leads to infrared divergences; these can
be controlled by a dynamical renormalization group for sufficiently
small $y$; for $y=4$ this yields a K41-type $k^{-5/3}$ spectrum
at the one-loop level. This value of $y$ is too large to trust a low-$y$,
one-loop result; also, for $y \geq 3$, the sweeping effect leads to
another singularity~\cite{mou}. Nevertheless, this randomly forced model
has played an important role historically. Thus it has been studied
numerically via the pseudo-spectral method~\cite{sainmanu,bifpower}.
These studies have shown that, even though the stochastic forcing destroys
the vorticity tubes that we have described above, it yields multiscaling
of velocity structure that is consistent, for $y=4$, with the analogous
multiscaling in the conventional 3D Navier-Stokes equation, barring
logarithmic corrections. We will discuss the analogue of this problem for
the stochastically forced Burgers equation in Sec. 6.4.

\subsection{Shell Models}
Even though shell models are far simpler than their parent partial
differential equations (PDEs), they cannot be solved analytically.  The
multiscaling of equal-time structure functions in such models has been
investigated numerically by several groups; 
an overview of earlier work and details about numerical methods for the
 stiff shell-model equations can be found in Refs.
~\cite{epjb,ssrnjp,shellrev}.
An
illustrative plot of equal-time multiscaling exponents for the GOY shell
model is given in the right panel of Fig. 1.

We devote the rest of this Subsection to a discussion of the dynamic
multiscaling of time-dependent shell-model structure functions that has been
elucidated recently by our group~\cite{epjb,ssrnjp,dmrpprl,dmpass}. So far,
detailed numerical studies of such dynamic multiscaling has been possible
only in shell models. We concentrate on time-dependent velocity
structure functions in the GOY model and their passive-scalar analogues
in the advection-diffusion shell model.

In a typical decaying-turbulence experiment or simulation, energy is injected
into the system at large length scales (small $k$); it then cascades to small
length scales (large $k$); eventually viscous losses set in when the energy
reaches the dissipation scale. We will refer to this as cascade completion.
Energy spectra and structure functions show power-law forms like their
counterparts in statistically steady turbulence.  It turns out~\cite{ssrnjp}
that the multiscaling exponents for both equal-time and time-dependent
structure functions are universal in so far as they are independent of
whether they are measured in decaying turbulence or the forced case in which
we get statistically steady turbulence.

Furthermore, the distinction between Eulerian and Lagrangian frameworks
assumes special importance in the study of dynamic multiscaling of
time-dependent structure functions. Eulerian-velocity structure functions are
dominated by the sweeping effect that lies at the heart of Taylor's
frozen-flow hypothesis; this relates spatial and temporal separations
linearly (see Sec. 2) whence we obtain trivial dynamic scaling with dynamic
exponents $z_p^{\cal{E}} = 1$ for all $p$, where the superscript $\cal{E}$
stands for Eulerian. By contrast, we expect nontrivial dynamic multiscaling
in Lagrangian or quasi-Lagrangian measurements. Such measurements are
daunting in both experiments and direct numerical simulations; however,
they are possible in shell models. As we have mentioned in Sec. 3, shell
models have a quasi-Lagrangian character since they do not have direct
sweeping effects. Thus we expect nontrivial dynamic multiscaling of
time-dependent structure functions in them.

Indeed, we find that~\cite{epjb,ssrnjp,lvov} that, given a time-dependent
structure function, we can extract an infinity of time scales from it.
Dynamic scaling Ans\"atze [cf., Eq. (4)] can then be used to extract dynamic
multiscaling exponents. A generalisation of the multifractal model then
suggests linear relations, referred to as bridge relations, between these
dynamic multiscaling exponents and their equal-time counterparts. These
can be related to equal-time exponents via bridge relations.
We show how to check these bridge relations in shell models. However,
before we present details, we must define time-dependent
structure functions precisely.

The order-$p$, time-dependent, structure functions, for
longitudinal velocity increments,
$\delta u_{\parallel}({\bf x},{\bf r},t) \equiv
         [{\bf u}({\bf x}+{\bf r},t) - {\bf u}({\bf x},t)]$
and passive-scalar increments,
$\delta\theta({\bf x},t,{\bf r}) = \theta({\bf x} + {\bf r},t) - \theta({\bf
x},t)$
are defined as
\begin{eqnarray}
{\mathcal F}_p^{u}(r,\{t_1,\ldots,t_p\}) \equiv
       \la [\delta u_{\parallel}({\bf x},t_1,r) \ldots
             \delta u_{\parallel}({\bf x},t_p,r)] \ra
\label{Fp}
\end{eqnarray}
and
\begin{equation}
{\mathcal F}_p^{\theta}({\bf r},{t_1,...,t_p}) = <[\delta\theta({\bf x},t_1,{\bf
r})...\delta\theta({\bf x},t_p,{\bf r})]>;
\label{Fp1}
\end{equation}
i.e., fluctuations are probed over a length scale $r$ which lies in the
inertial range. For simplicity, we consider $t_1=t$ and $t_2=\ldots=t_p=0$ in
both Eq. (\ref{Fp}) and Eq. (\ref{Fp1}).  Given 
${\mathcal F}^{u}(r,t)$ and
${\mathcal F}^{\theta}(r,t)$,  we can define the order-$p$, degree-$M$,
integral-time scales and derivative-time scales as follows~\cite{ssrnjp}:
\begin{equation}
{\cal T}^{I,u}_{p,M}(r,t) \equiv
 \biggl[ \frac{1}{{\mathcal S}^{u}_p(r)}
\int_0^{\infty}{\mathcal F}^{u}_p(r,t)t^{(M-1)} dt
\biggl]^{(1/M)};
\label{timp}
\end{equation}
\begin{equation}
{\cal T}^{I,\theta}_{p,M}(r,t) \equiv
 \biggl[ \frac{1}{{\mathcal S}^{\theta}_p(r)}
\int_0^{\infty}{\mathcal F}^{\theta}_p(r,t)t^{(M-1)} dt
\biggl]^{(1/M)};
\label{timp1}
\end{equation}

\begin{equation}
{\cal T}^{D,u}_{p,M}(r,t) \equiv
 \biggl[ \frac{1}{{\mathcal S}^{u}_p(r)}
\frac{\partial^M {\mathcal F}^{u}_p(r,t)}{\partial t^M}
\biggl]^{(-1/M)};
\label{tdp}
\end{equation}
\begin{equation}
{\cal T}^{D,\theta}_{p,M}(r,t) \equiv
 \biggl[ \frac{1}{{\mathcal S}^{\theta}_p(r)}
\frac{\partial^M {\mathcal F}^{\theta}_p(r,t)}{\partial t^M}
\biggl]^{(-1/M)}.
\label{tdp1}
\end{equation}

Integral-time dynamic multiscaling exponents $z^{I,u}_{p,M}$
for fluid turbulence can be defined via
${\cal T}^{I,u}_{p,M}(r,t) \sim r^{z^{I,u}_{p,M}}$
and the derivative-time ones $z^{D,u}_{p,M}$ by
${\cal T}^{D,u}_{p,M}(r,t) \sim r^{z^{D,u}_{p,M}}$.
They satisfy the following bridge relations~\cite{ssrnjp}:
\begin{equation}
z^{I,u}_{p,M} = 1 + [\zeta_{p-M} - \zeta_p]/M;
\label{zipm}
\end{equation}
\begin{equation}
z^{D,u}_{p,M} = 1 + [\zeta_p - \zeta_{p+M}]/M.
\label{zdpm}
\end{equation}
For passive-scalars advected by a turbulent velocity field, the
corresponding dynamic multiscaling exponents are defined as
${\cal T}^{I,\theta}_{p,M}(r,t)
\propto r^{z^{I,\theta}_{p,M}}$ and ${\cal T}^{D,\theta}_{p,M}(r,t)
\propto r^{z^{D,\theta}_{p,M}}$; they satisfy the following bridge
relations involving the scaling exponents $\zeta_M$ of equal-time,
order-$M$ structure functions of the advecting velocity field:
\begin{eqnarray}
 z^{I,\theta}_{p,M} = 1 - \frac{\zeta_M}{M}, \hspace{1cm} z^{D,\theta}_{p,M} = 1
- \frac{\zeta_{-M}}{M}.
 \label{}
\end{eqnarray}
These bridge relations, unlike Eq. (\ref{zipm}) and Eq. (\ref{zdpm}),
 are independent of $p$. [Recall that, for the
Kraichnan model, we have already shown in Sec. 5 that we get simple
dynamic scaling.]

GOY-model equal-time structure functions and their associated inertial-range
exponents are defined as follows:
\begin{equation}
S^{u}_p(k_n) \equiv \la [u_n(t)u^{\ast}_n(t)]^{p/2} \ra \sim k_n^{-\zeta_p}.
\label{goy1}
\end{equation}
The time-dependent structure function are
\begin{equation}
F^{u}_p(k_n,t_0,t) \equiv \la [u_n(t_0)u^{\ast}_n(t_0 + t)]^{p/2} \ra.
\label{goy2}
\end{equation}
We evaluate these numerically for the GOY shell model [numerical details may
be found in Refs.~\cite{epjb,ssrnjp}], extract integral and derivative time
scales from them and thence the exponents $z^{I,u}_{p,1}$ and
$z^{D,u}_{p,2}$, respectively, from slopes of  log-log plots of
$T^{I,u}_{p,1}(n)$  versus $k_n$ (right panel of Fig. 6) and of $T^{D,u}_{p,2}(n)$
versus $k_n$ (right panel of Fig. 7).
\begin{figure}[htbp]
\epsfxsize=6cm
\centerline{\epsfbox{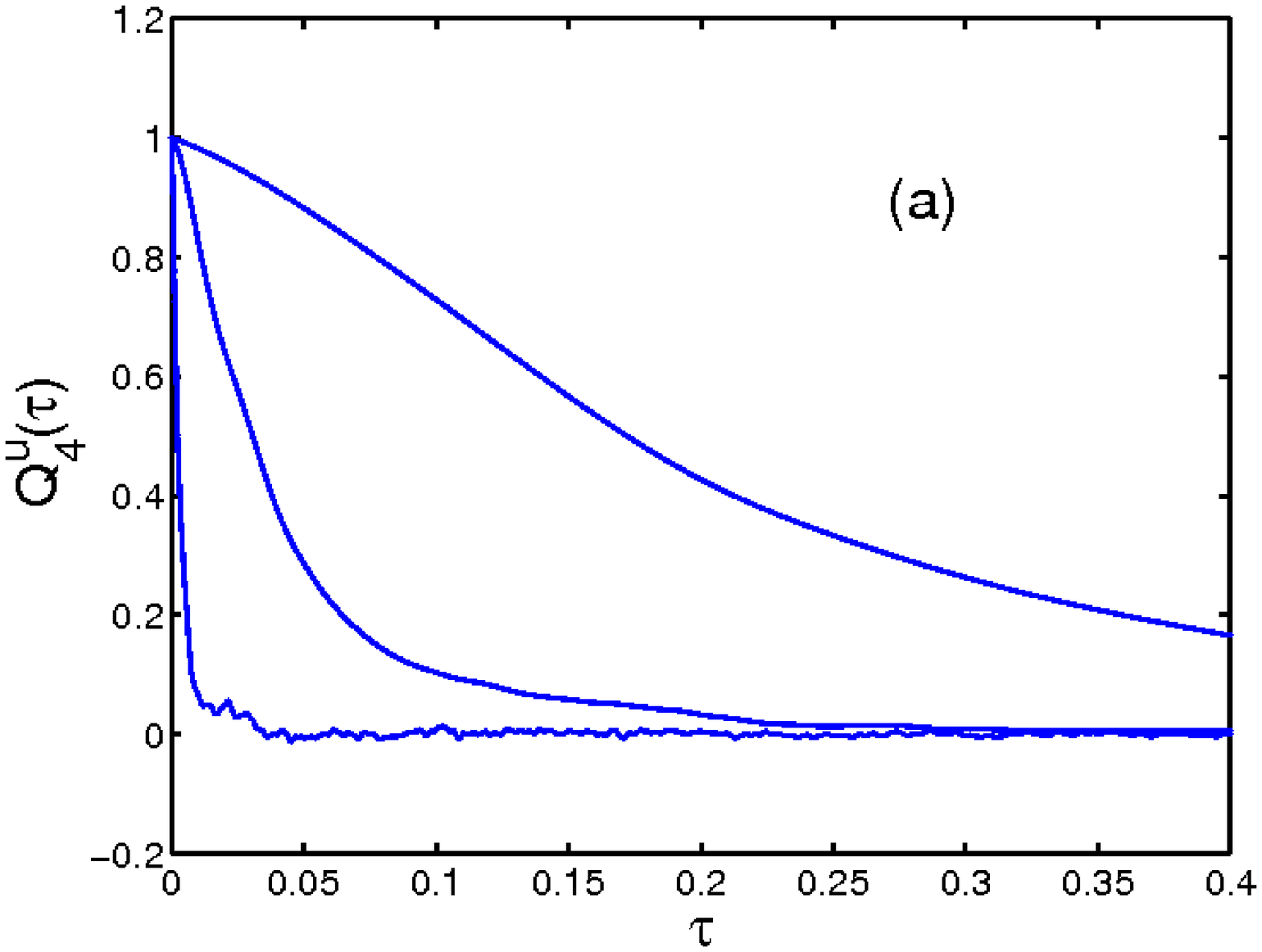},\epsfxsize=6cm
\epsfbox{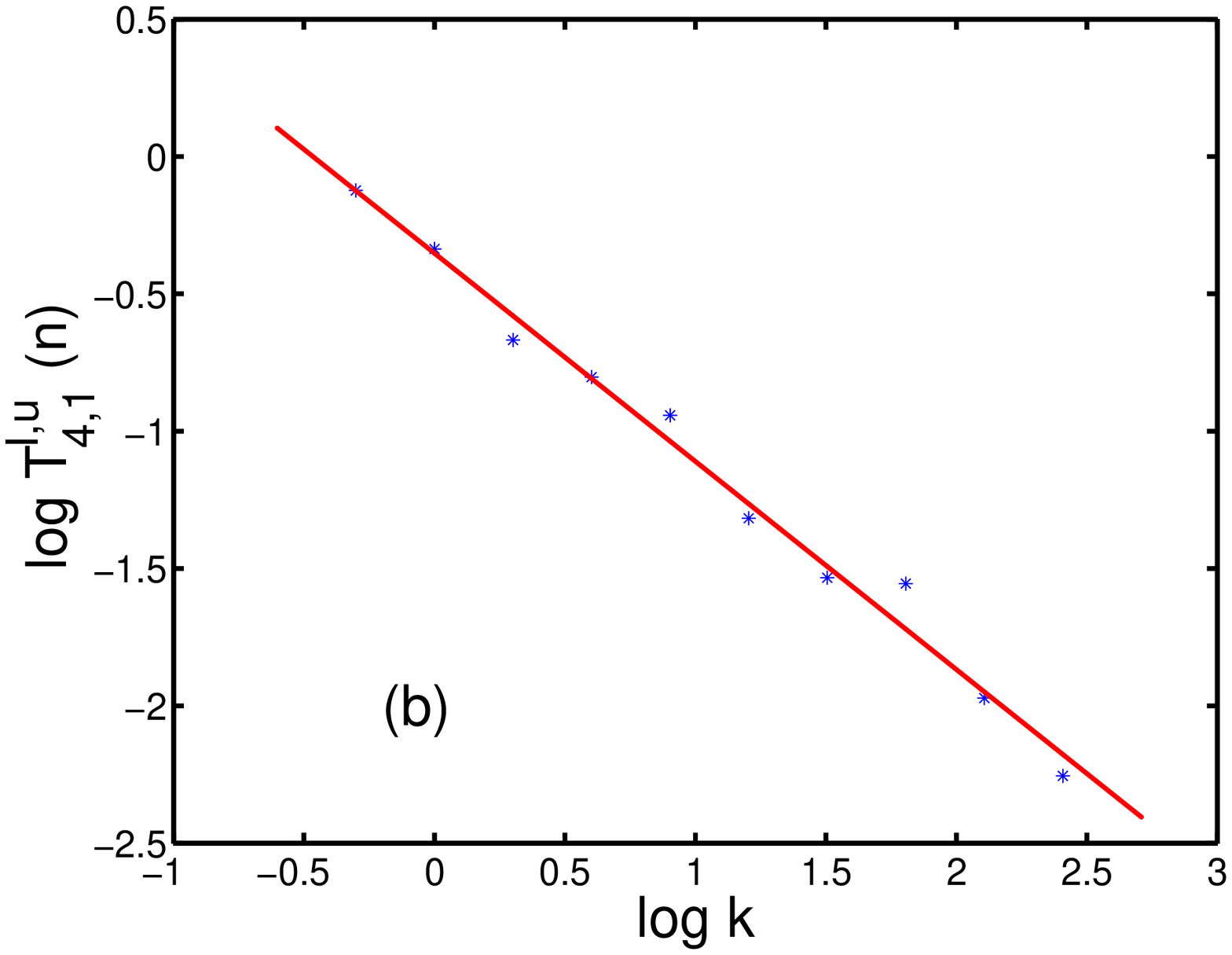}}
\caption{(Color online) (a) A  representative plot of the normalised fourth order
time-dependent
structure function versus the dimensionless time $\tau$ obtained
from the GOY shell model. The plots are for shells 4, 6, and 8
(from top to bottom). (b) A log-log plot of $T^{I,u}_{4,1}(n)$ 
versus $k$ (for convenience, we have dropped the subscript $n$
in the label of the x-axis in the figure); a linear fit gives the 
dynamic 
mulstiscaling exponent $z^{I,u}_{4,1}$.}
\label{z6}
\end{figure}

\begin{figure}[htbp]
\epsfxsize=6cm
\centerline{\epsfbox{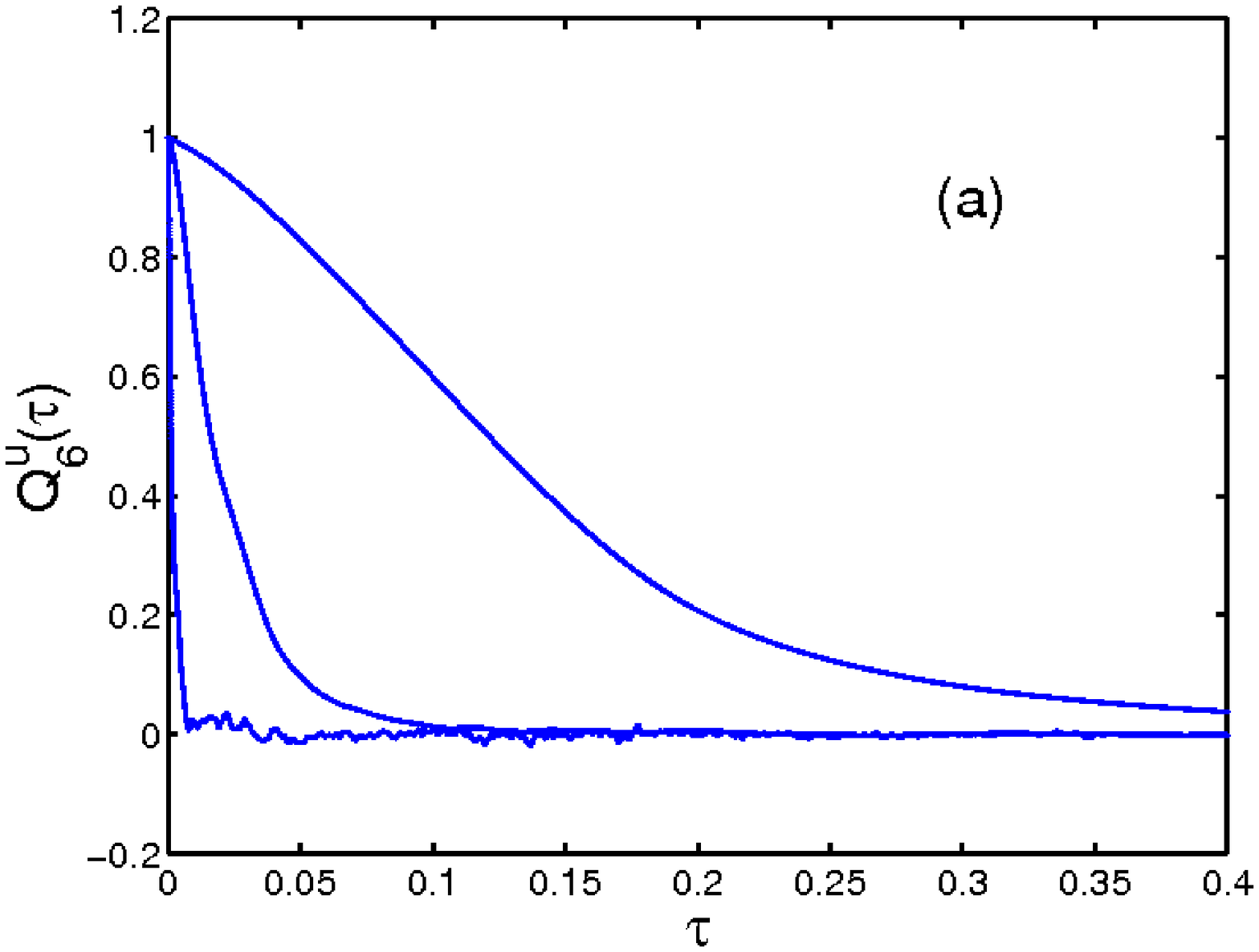},\epsfxsize=6cm
\epsfbox{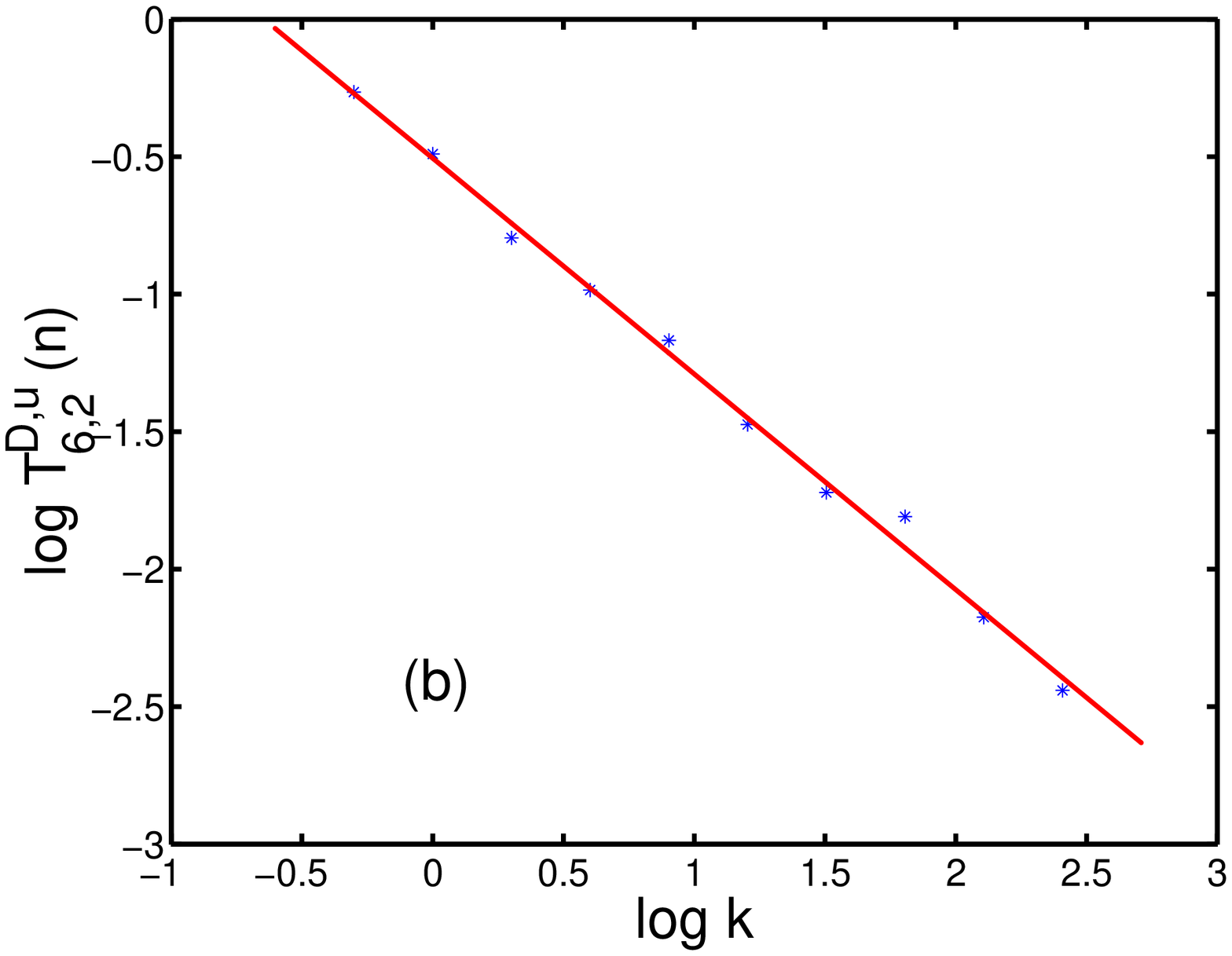}}
\caption{(Color online) (a) A  representative plot of the normalised sixth order
time-dependent
structure function versus the dimensionless time $\tau$ obtained
from the GOY shell model. The plots are for shells 4, 6, and 8
(from top to bottom). (b) A log-log plot of $T^{D,u}_{6,2}(n)$ 
versus $k$ (for convenience, we have dropped the subscript $n$
in the label of the x-axis in the figure); a linear fit gives the dynamic multiscaling exponent
$z^{D,u}_{6,2}$.}
\label{z6}
\end{figure}

There is excellent agreement (within error bars) of the multiscaling
exponents $z^{I,u}_{p,1}$ and $z^{D,u}_{p,2}$, obtained from our simulations,
with the values computed from the appropriate bridge relations using the
equal-time exponents, $\zeta_p$. 

For the passive-scalar case, the equal-time order-$p$ structure functions
is
\begin{equation}
S^{\theta}_p(k_n) \equiv \la [\theta(t)\theta^{\ast}_n(t)]^{p/2} \ra \sim
k_n^{-\zeta^{\theta}_p}
\label{goy1}
\end{equation}
and its time-dependent version is
\begin{equation}
F^{\theta}_p(k_n,t_0,t) = <[\theta_n(t_0)\theta_n^*(t_0 + t)]^{p/2}>.
\label{}
\end{equation}
We consider decaying turbulence here with $t_0$ a time origin.
It is useful now to work with the normalised
time-dependent structure function, $Q^{\theta}_p(n,t_0,t) =
\frac {F^{\theta}_p(k_n,t_0,t)}{F^{\theta}_p(k_n,t,0)}$.
For the case of passive-scalars advected by a velocity field which
is turbulent (a solution of the GOY model), we calculate the
integral (for $M$ = $1$) and derivative time scales (for $M$ = $2$)
corresponding to Eq.(58) and Eq.(60),
respectively. The slope of a log-log plot of $T^{I,\theta}_{p,1}(n)$
 {\it vs} $k_n$ yields the integral time scale
exponent, $z^{I,\theta}_{p,1}$, since $T^{I,\theta}_{p,1}(n)
\propto k_n^{-z^{I,\theta}_{p,1}}$. Likewise, from plots of the
derivative time scales we extract the exponent $z^{D,\theta}_{p,2}$.
For a detailed discussion on dynamic multiscaling in this model
we refer the reader to Refs.~\cite{ssrnjp,dmrpprl}.

\subsection{2D Navier-Stokes Turbulence}

We now consider illustrative numerical calculations for the 2D NS
equations (9)-(11). We begin with periodic boundary conditions for
which we can use a pseudo-spectral method similar to the one given 
in the
previous Subsection for the 3D NS case. We study decaying turbulence
first with the source function $f$ (the ${\hat{\bf z}}$ component of
the curl of some force ${\nabla \times \bf F}$) set to 0.
We use $1024^2$ collocation points and the standard $2/3$ dealiasing
procedure; for time marching we use a second-order Runge-Kutta
scheme~\cite{nr}. Our initial condition $|\omega(k)|^2 = k^{-3}
\exp(-k^2)$ leads to a forward cascade. We seed the flow with
Lagrangian tracers and use a cubic spline interpolation method to
calculate their trajectories ~\cite{yeungrev}. Representative 
plots from our from our
DNS are shown in Fig. 8. The first part (Fig. 8a) shows a
compensated energy spectrum $k^3 E(k)$ for the case with no Ekman
friction. Figure 8b, from a DNS with Ekman friction $\alpha_E =
0.1$, Kolmogorov forcing~\cite{prasad02}, and periodic boundary
conditions, shows a trajectory of a Lagrangian tracer superimposed
on a pseudocolour plot of the vorticity field at time $t=100$; the
tracer starts at the point marked with a circle ($t=0$) and ends at
the star ($t=100$). For a state-of-the-art simulation that resolves
both forward and inverse cascades in a forced DNS of 2D turbulence
we refer the reader to Ref.~\cite{bof07}; such DNS studies have also
investigated the scaling properties of structure functions and have
provided some evidence for conformal invariance in the inverse
cascade inertial range~\cite{conf}.

\begin{figure}[htbp]
\epsfxsize=6cm
\centerline{\epsfbox{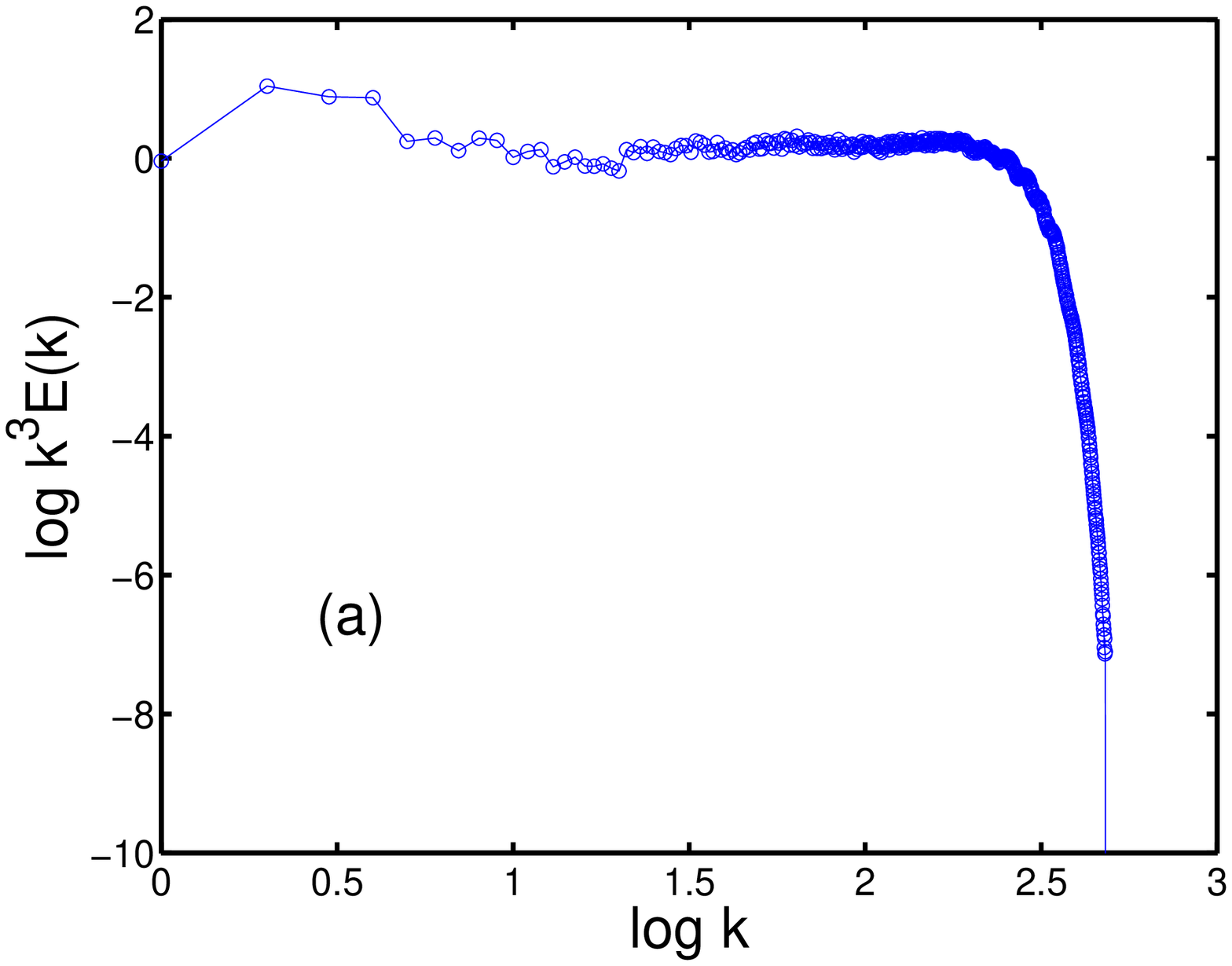},\epsfxsize=6cm
\epsfbox{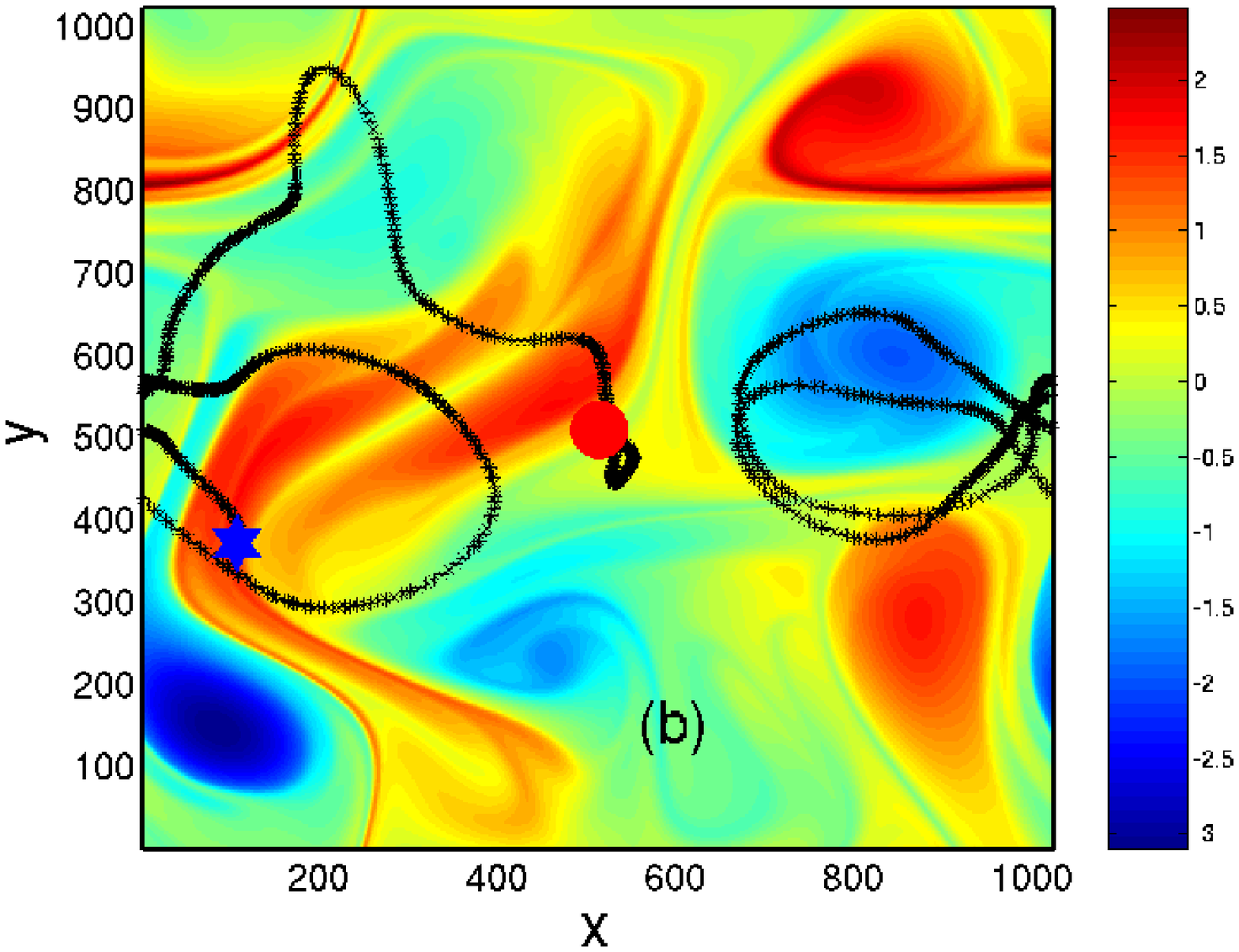}}
\caption{(Color online) (a) A log-log plot of the compensated energy spectrum 
$k^3E(k)$ versus $k$ from our DNS, of resolution $1024^2$, of
two dimensional decaying turbulence with periodic boundary
conditions. The flat region indicates a scaling form $E(k) \sim k^{-3}$.
(b) The trajectory of a single Lagrangian particle over a time of
order 100 in a two-dimensional flow with drag and forcing. The
starting point of the trajectory is in the middle of the box and is
indicated by a red circle; the end point is indicated
by a blue star. The trajectory is superimposed on
a pseudocolor plot of the vorticity field corresponding to the
time at the end of the Lagrangian trajectory. The figure 
corresponds to a forced 
DNS of resolution $1024^2$ with periodic boundary conditions, 
statistical steady state, and with a coefficient of 
Ekman friction $\alpha_E = 0.1$.}
\label{2DEk}
\end{figure}

We end with an illustrative example of a recent DNS
study~\cite{prasad02} that sheds light on the effect of the Ekman
friction on the statistics of the forward cascade in wall-bounded
flows that are directly relevant  to laboratory soap-film
experiments~\cite{riv00,riv01,dan02,riv07}. 
The details of this DNS are given in Ref.~\cite{prasad02}.  In
brief, $\omega$ is driven to a statistical steady state by a
deterministic Kolmogorov forcing $F_{\omega} \equiv k_{inj} F_0
\cos(k_{inj}x)$, with $F_0$ the amplitude and $k_{inj}$ the
wavenumber on which the force acts; no-slip and no-penetration
boundary conditions are imposed on the walls. The important
non-dimensional control parameters are the Grashof number ${\mathcal
G}=2\pi||F_\omega||_2/(k_{inj}^{3}\rho\nu^2)$ and the
non-dimensional Ekman friction $\gamma=\alpha_E/ (k_{inj}^2\nu)$,
where we non-dimensionalize $F_\omega$ by $2\pi/(k_{inj} ||
F_\omega||_2)$, with $||F_\omega||_2 \equiv (\int_{A}
|F_\omega|^2d{\bf x})^{1/2}$ and the length and time scales are made
non-dimensional by scaling ${\bf x}$ by
$k_{inj}^{-1}$ and $t$ by $k_{inj}^{-2}/\nu$.  We use a fourth-order
Runge-Kutta scheme  for time marching and evaluate spatial
derivatives via second-order and fourth-order, centered, finite
differences, respectively, for points adjacent to the walls and for
points inside the domain. The Poisson equation is solved by using a
fast-Poisson solver~\cite{nr} and $\omega$ is calculated at the
boundaries by using Thom's formula~\cite{prasad02}. 

Since Kolmogorov forcing is inhomogeneous, we use the decomposition
$\psi = \langle \psi\rangle + \psi'$ and $\omega = \langle \omega
\rangle + \omega'$, where the angular brackets denote a time average
and the prime the fluctuating part to calculate the order-$p$
velocity and vorticity structure functions. Since this is a
wall-bounded flow, it is important to extract the isotropic parts of
these structure functions~\cite{prasad02,anisrev}. Furthermore,
given our resolution ($2049^2$), it becomes necessary to use the 
ESS procedure to extract exponent ratios. Illustrative log-log
ESS plots for velocity, $S_p(R)$, and vorticity, $S^{\omega}_p(R)$, 
structure functions are shown in the left and right panels,
respectively, of Fig. 9; their slopes yield the exponent ratios
that are plotted versus the order $p$ in Fig. 10.
The Kraichnan-Leith-Batchelor (KLB) predictions~\cite{kraic67} for
these exponent ratios, namely, $\zeta^{KLB}_p/\zeta^{KLB}_2 \sim
r^{p/2}$ and $\zeta_p^{\omega,KLB}/\zeta_2^{\omega,KLB} \sim r^0$,
agree with our values for $\zeta_p/\zeta_2$ but not
$\zeta_p^\omega/\zeta_2^\omega$: velocity structure functions do not
display multiscaling~[left panel of Fig. 10] whereas their
vorticity analogs do~[note the curvature of the plot in the right
panel of Fig. 10].  Similar results have been seen in DNS
studies with periodic boundary conditions~\cite{tsa05,bof07}. 
Additional results for PDFs of several properties can
be obtained from our DNS~\cite{prasad02}; these are in striking
agreement with experimental results~\cite{riv01}.  

\begin{figure}[htbp]
\epsfxsize=6cm
\centerline{\epsfbox{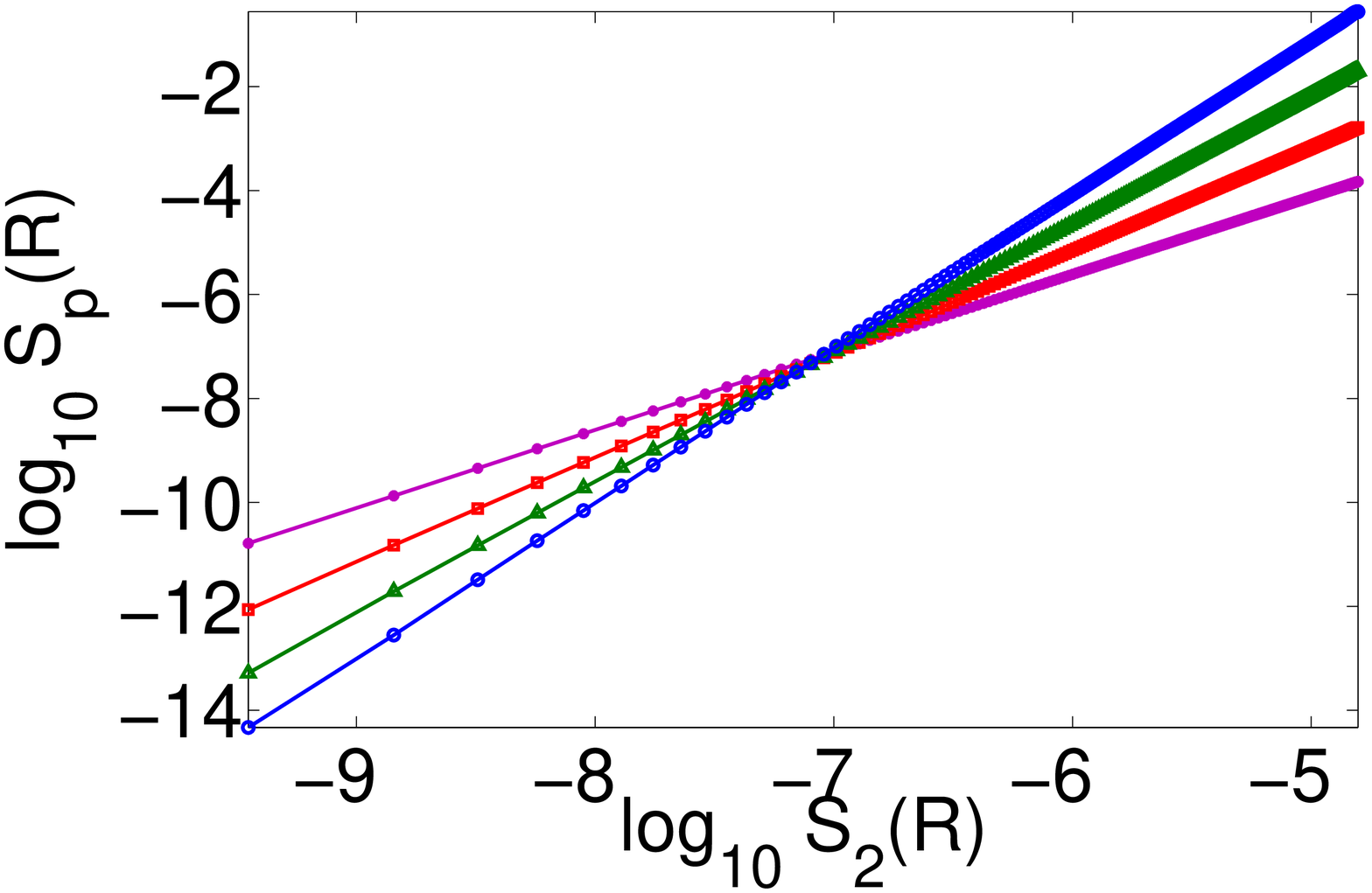} 
\epsfxsize=6cm \epsfbox{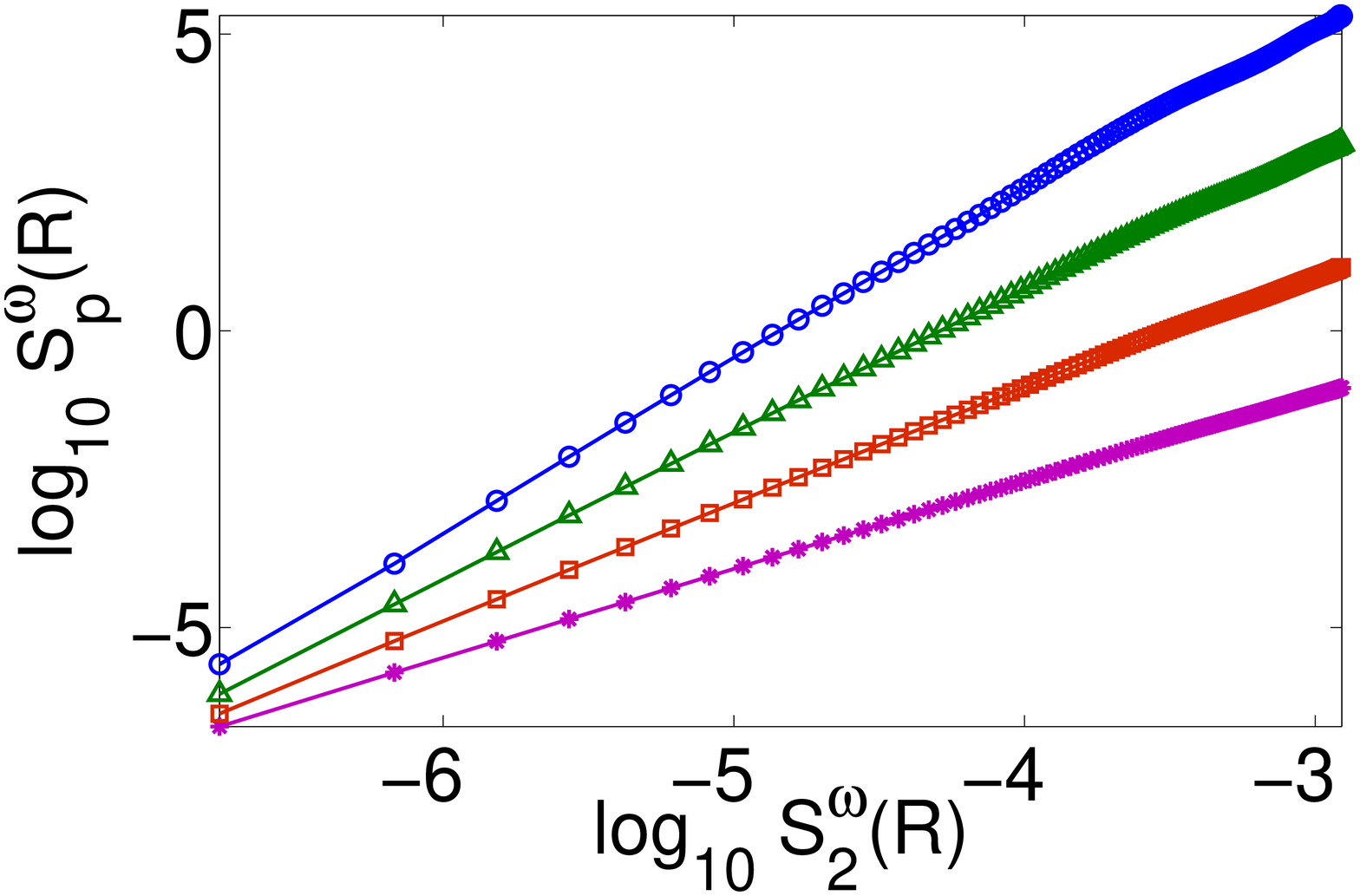}} 
\caption{(Color online) (Left) Log-log ESS plots of the isotropic 
parts of the order-$p$  velocity structure functions $S_p(R)$ versus 
$S_2(R)$; $p=3$ (purple line with dots), $p=4$ 
(red line with square), $p=5$ (green line with triangles), and 
$p=6$ (blue line with circles). According to the KLB prediction 
$S_p(R) \sim R^\zeta_p$. (Right) Log-log ESS plots of the isotropic 
parts of the order-$p$ vorticity structure functions $S_p(R)$ 
versus $S_2(R)$; $p=3$ (purple line with stars), $p=4$ (red line 
with square), $p=5$ (green line with triangles), and $p=6$ (blue 
line with circles).}
\label{3diso1}
\end{figure}

\begin{figure}[htbp]
\epsfxsize=6cm
\centerline{\epsfbox{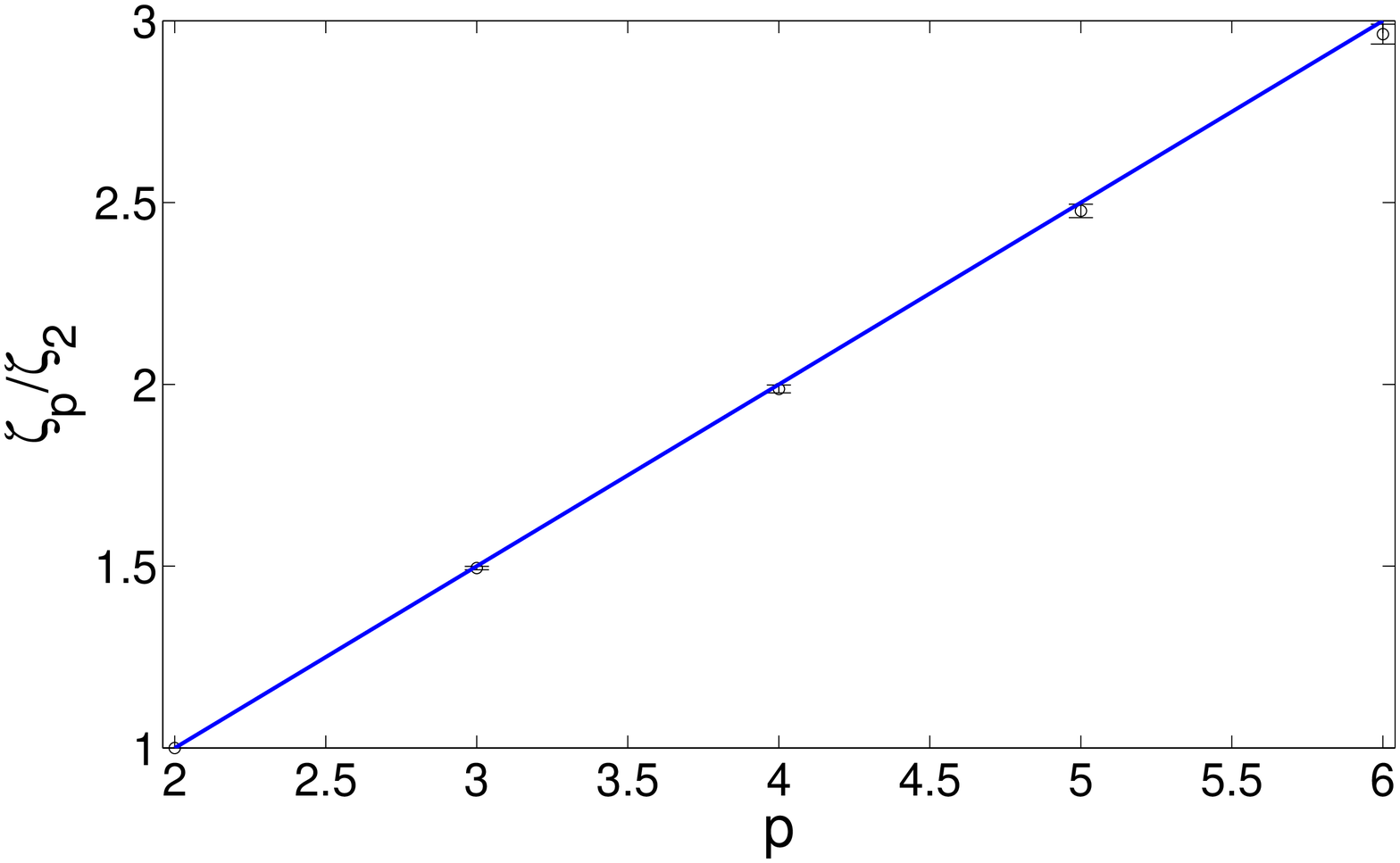} 
\epsfxsize=5.4cm \epsfbox{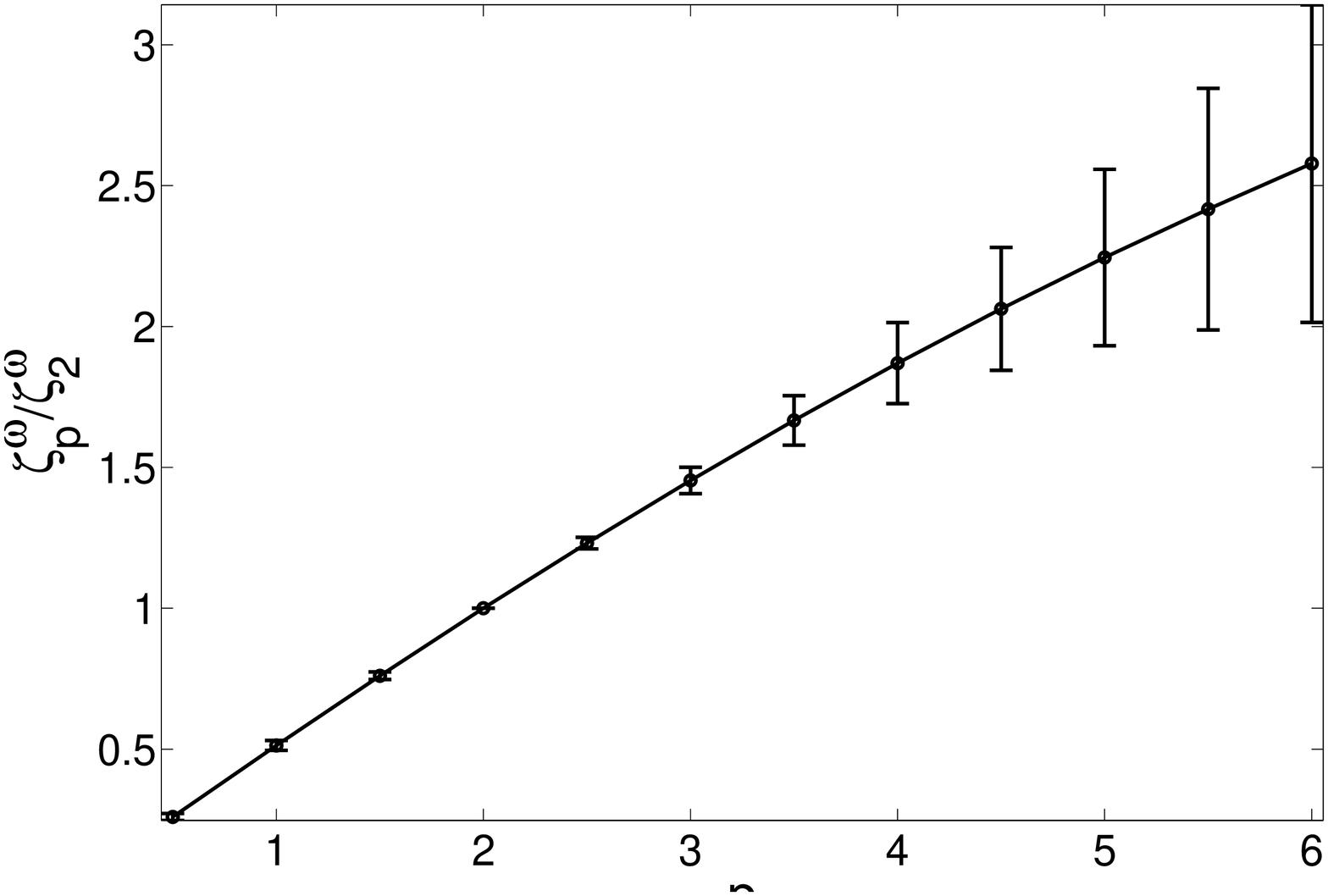}} 
\caption{(Color online) (Left) Plots of the exponent ratios 
$\zeta_p/\zeta_2$ versus $p$ for the velocity differences.
(Right) Plots of the exponent ratios $\zeta^\omega_p/\zeta^\omega_2$
 versus $p$ for the vorticity differences.}
\label{3diso1}
\end{figure}


\subsection{The One dimensional Burgers Equation}
In this Subsection we present a few representative numerical
studies of the 1D Burgers equation. The first of these uses a 
pseudo-spectral method with $2^{14}$ collocation points, the
$2/3$ dealising rule, and a fourth-order Runge-Kutta time-marching
scheme. In the second study of a stochastically forced Burgers
equation (see below) we use a fast-Legendre method that yields
results in the zero-viscosity limit~\cite{dmburg}.

For the Burgers equation with no external forcing and sufficiently
well-behaved initial conditions, the velocity field develops {\it
shocks}, or jump discontinuities, which merge into each other with
time. The time at which the first shock appears is usually denoted
by $t_*$. For all times greater than $t_*$, it is possible to
calculate, analytically, the scaling exponents $\zeta_p$ for the
equal-time structure functions via $S_p\equiv \langle
[u(x+r,t)-u(x)]^p\rangle \sim C_p |r|^p + C_p'|r|$, where the first
term comes from the {\it ramp}, and the second term comes from the
probability of having a shock in the interval $|r|$. As a
consequence of this we have {\it bifractal} scaling : for $0<p<1$
the first term dominates leading to $\zeta_p = p$ and for $p>1$ the
second one dominates giving $\zeta_p=1$. This leads to an energy
spectrum $E(k) \sim k^{-2}$. Representative plots from our 
pseudo-spectral DNS, with $\nu = 10^{-3}$ and an initial condition
$u(x) =$sin($x$) (for which $t_* = 1$) are shown in 
Fig. 11; the left panel shows plots of the velocity field at 
times $t = 0, \, 1,$ and $t = 1.5$ and the right panel the energy 
spectrum at $t = 1$.

\begin{figure}[htbp]
\epsfxsize=6cm
\centerline{\epsfbox{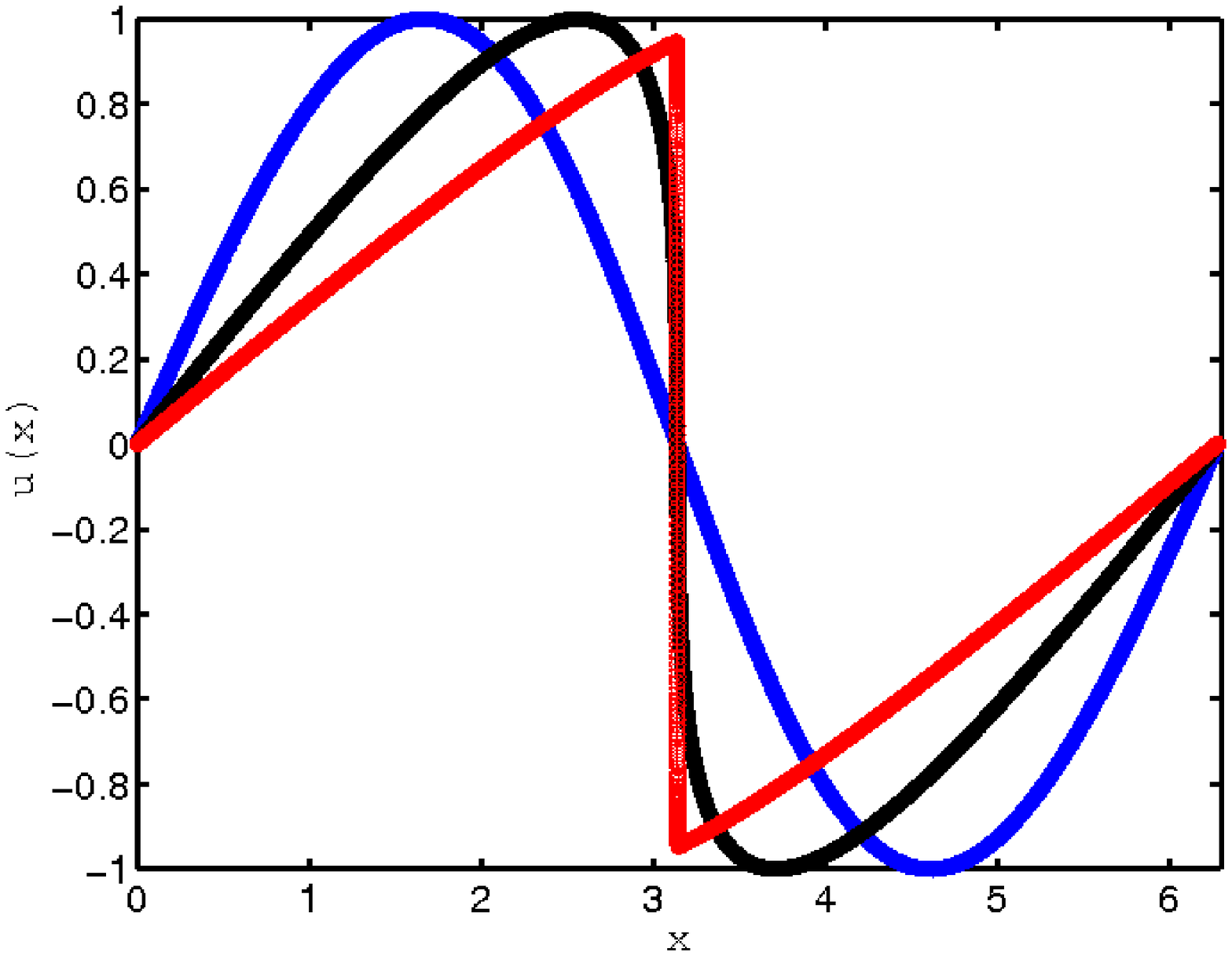},\epsfxsize=6cm
\epsfbox{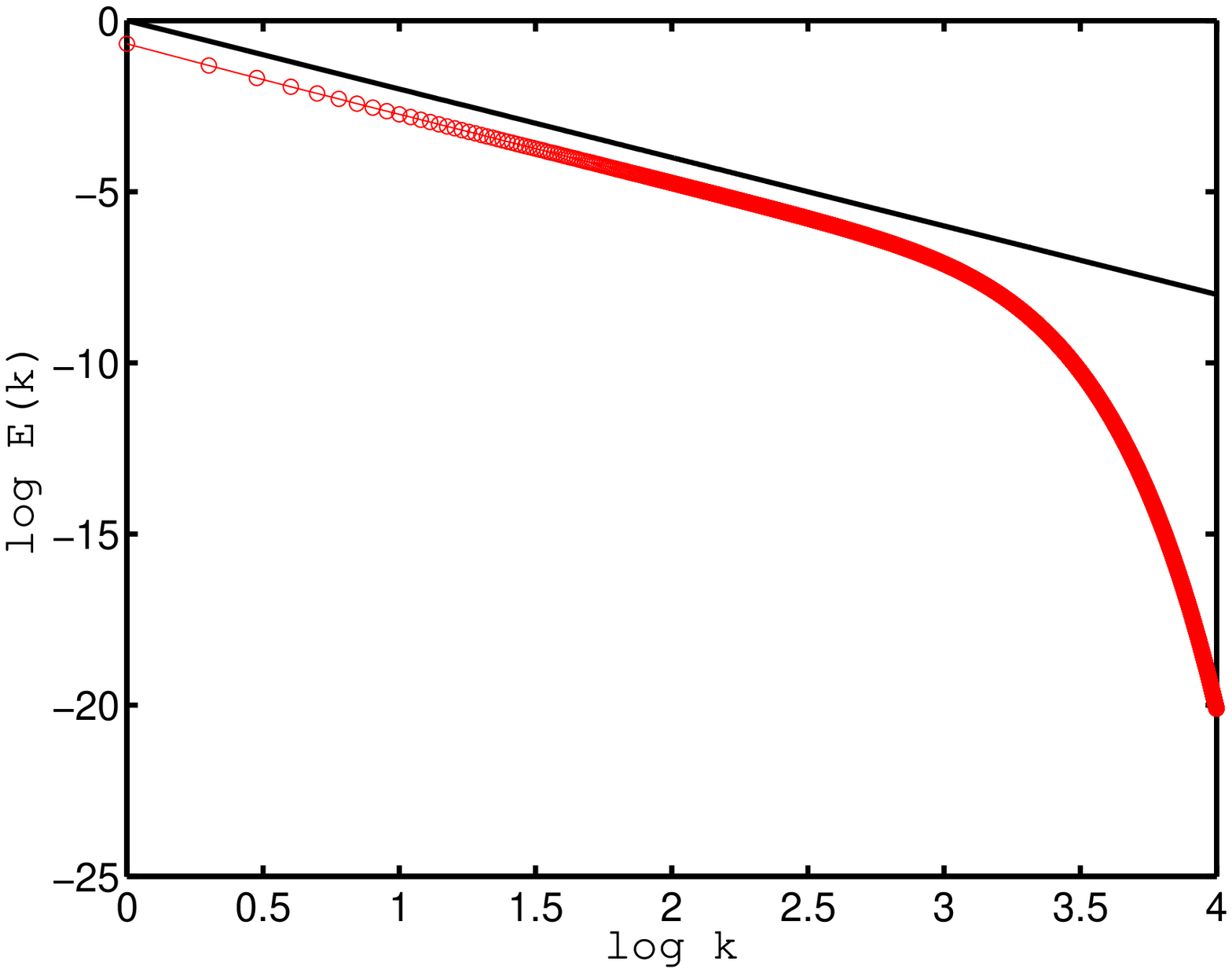}}
\caption{(Color online) (Left) Snapshots of the solution of 
the Burgers equation obtained from our DNS with initial condition 
$u(x) = $sin $x$ at times $t=0$ (blue), $t=1$ (black) and $t=2$ 
(red). (Right) A representative log-log plot of $E(k)$ versus 
$k$, at time $t=1$ for the Burgers equation with initial conditions 
$u(x) =$ sin $x$.}
\label{simpleburg}
\end{figure}

The stochastically forced Burgers equation has played an important
role in renormalization-group studies~\cite{dmburg}. 
In particular, consider a Gaussian random force  
$f(x,t)$ with zero mean and the following covariance in Fourier
space:
\begin{eqnarray}
\langle{\hat f}(k_1,t_1){\hat f}(k_2,t_2)\rangle =
        2D_0|k|^{\beta}\delta(t_1-t_2)\delta(k_1+k_2);  
\label{eq:force}
\end{eqnarray}
here ${\hat f}(k,t)$ is the spatial Fourier transform of $f(x,t)$,
$D_0$ is a constant, and the scaling properties of the forcing is
governed by the exponent $\beta$. For positive values of $\beta$,
the Burgers equation can be studied by using renormalization-group
techniques; specifically, for $\beta = 2$ one recovers simple 
(Kardar-Parisi-Zhang or KPZ)
scaling with the equal-time exponent $\zeta_p = p$.  It was hoped
that  forcing with negative values of $\beta$ (in particular $\beta
= -1$), which cannot be studied by renormalization-group methods,
might yield multiscaling of velocity structure functions. 

However, our high-resolution study~\cite{dmburg}, which uses a
fast-Legendre method, has shown that the apparent multiscaling of
structure functions in this stochastic model might arise because of
numerical artifacts. The general consensus is that this
stochastically forced Burgers model should show bifractal scaling.
In Fig. 12 we present representative plots of the velocity field
(left panel, blue curve) and the scaling exponents (right panel) for
this model.  We have obtained the data for these figures by using a
fast-Legendre method with $2^{18}$ collocation points.  

\begin{figure}[htbp]
\epsfxsize=6cm
\centerline{\epsfbox{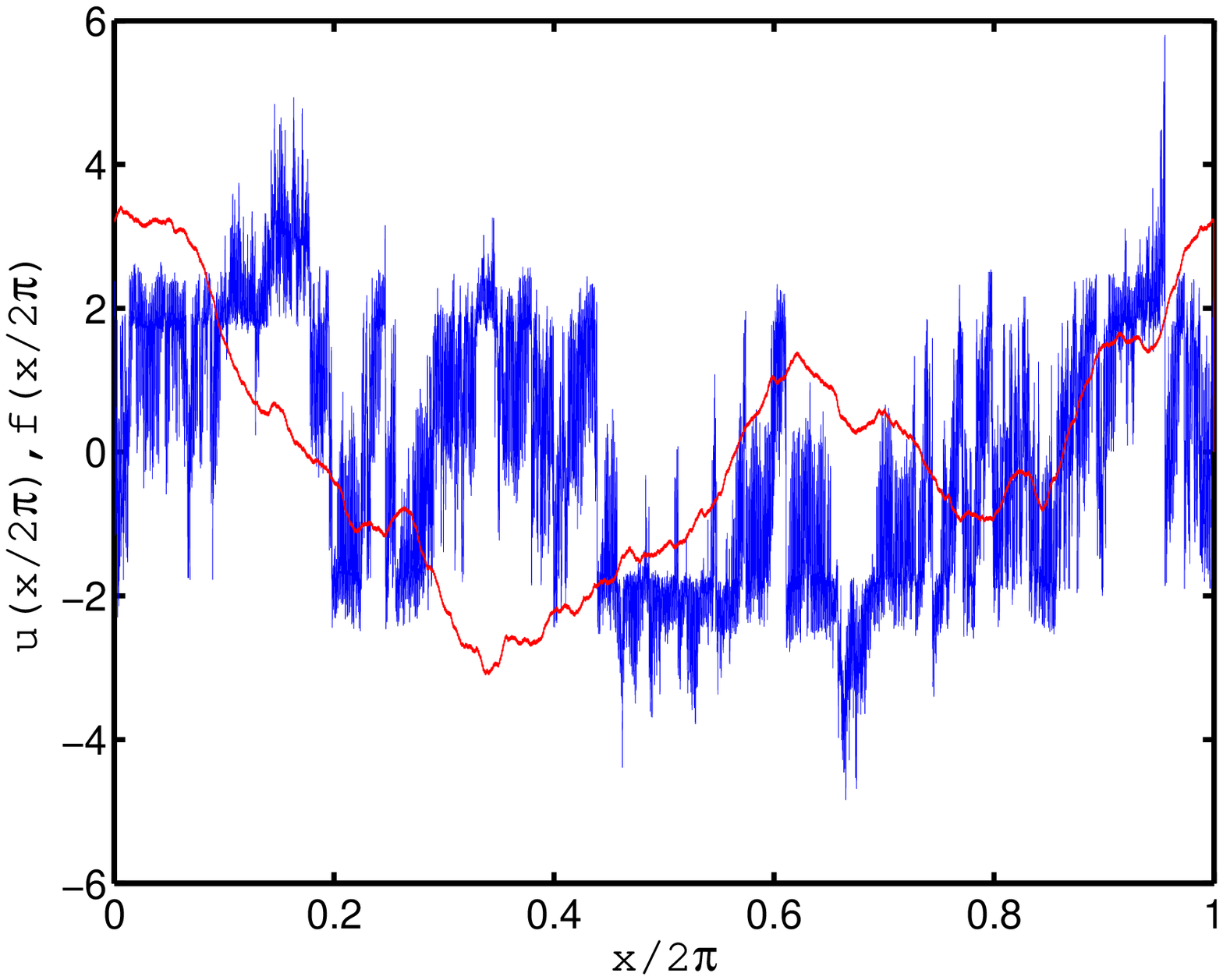},\epsfxsize=6cm
\epsfbox{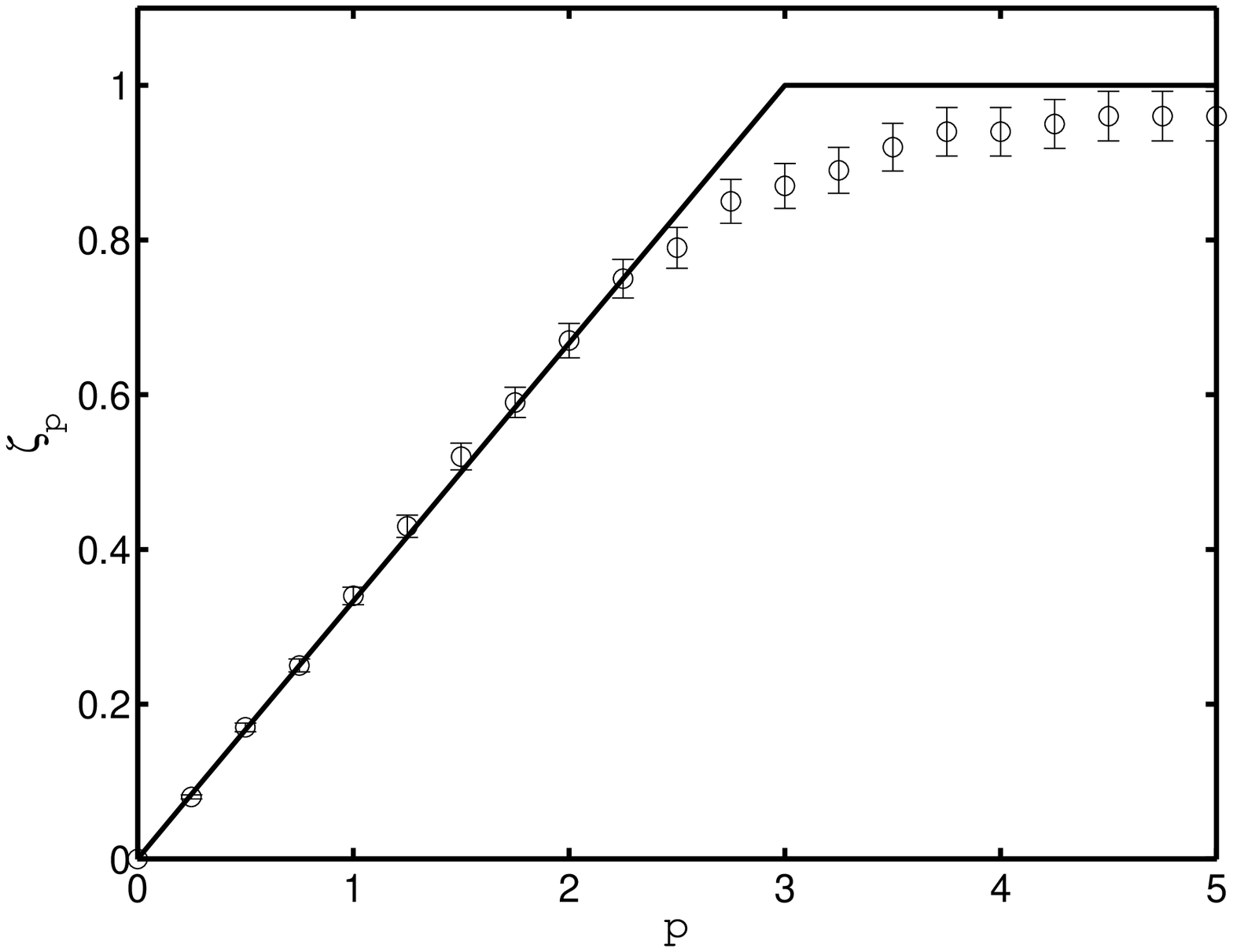}}
\caption{(Color online) (Left) A snapshot of the velocity field (jagged line in blue) in steady state and the force in red from our 
fast-Legendre method DNS of the stochastically forced Burgers 
equation. 
(Right) A representative plot of the exponents $\zeta_p$, with 
error-bars, for the equal-time velocity structure functions of the 
stochastically forced Burgers 
equation; bifractal scaling is shown by the black solid line; 
the deviations from this are believed to arise from artefacts (see 
text).} 
\label{stochburg}
\end{figure}

Numerical studies of the Burgers equation have also proved useful in
elucidating bottleneck structures in energy spectra 
~\cite{bottlesimul,bottleexper}(cf., the
spectrum in the left panel of Fig. 5). It turns out that such a
bottleneck does not occur in the conventional Burgers equation.
However, it does~\cite{botfrisch} occur in the hyperviscous one, in
which usual Laplacian dissipation operator is replaced by its
$\alpha^{\rm th}$ power; this is known as hyperviscosity for $\alpha
> 1$. We show a representative compensated energy spectrum for the
case $\alpha = 4$ in the left panel of Fig. 13.  We have obtained
this from a pseudo-spectral DNS with $2^{12}$ collocation points.  The
$\alpha \to \infty$ limit is very interesting too since, in this
limit, the hyperviscous Burgers equation maps on to the
Galerkin-truncated version of the inviscid Burgers equation.  In
this Galerkin-truncated inviscid case, the Fourier modes thermalise
~\cite{Lee52,KrAbEq};
in a compensated energy spectrum this shows up as $E(k) \sim k^2$,
for large $k$ [see the right panel of Fig. 13 for the case $\alpha =
200$].  Such thermalisation effects in the Galerkin-truncated Euler
equation have also attracted a lot of attention~\cite{Cichowlas};
and the link between bottlenecks and thermalisation has been
explored in our recent work~\cite{botfrisch} to which we refer the
interested reader.

\begin{figure}[htbp]
\epsfxsize=6cm
\centerline{\epsfbox{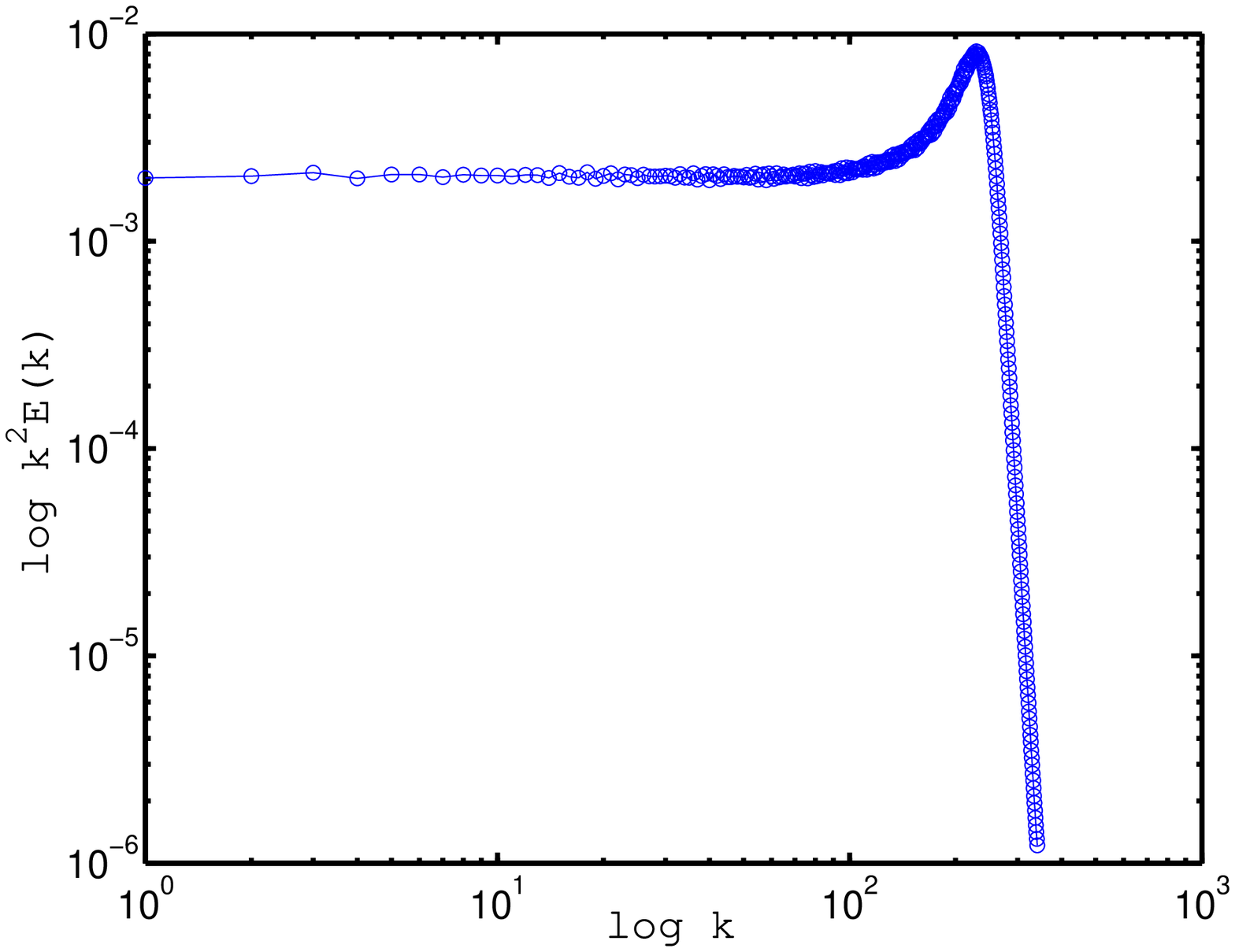},\epsfxsize=7.5cm
\epsfbox{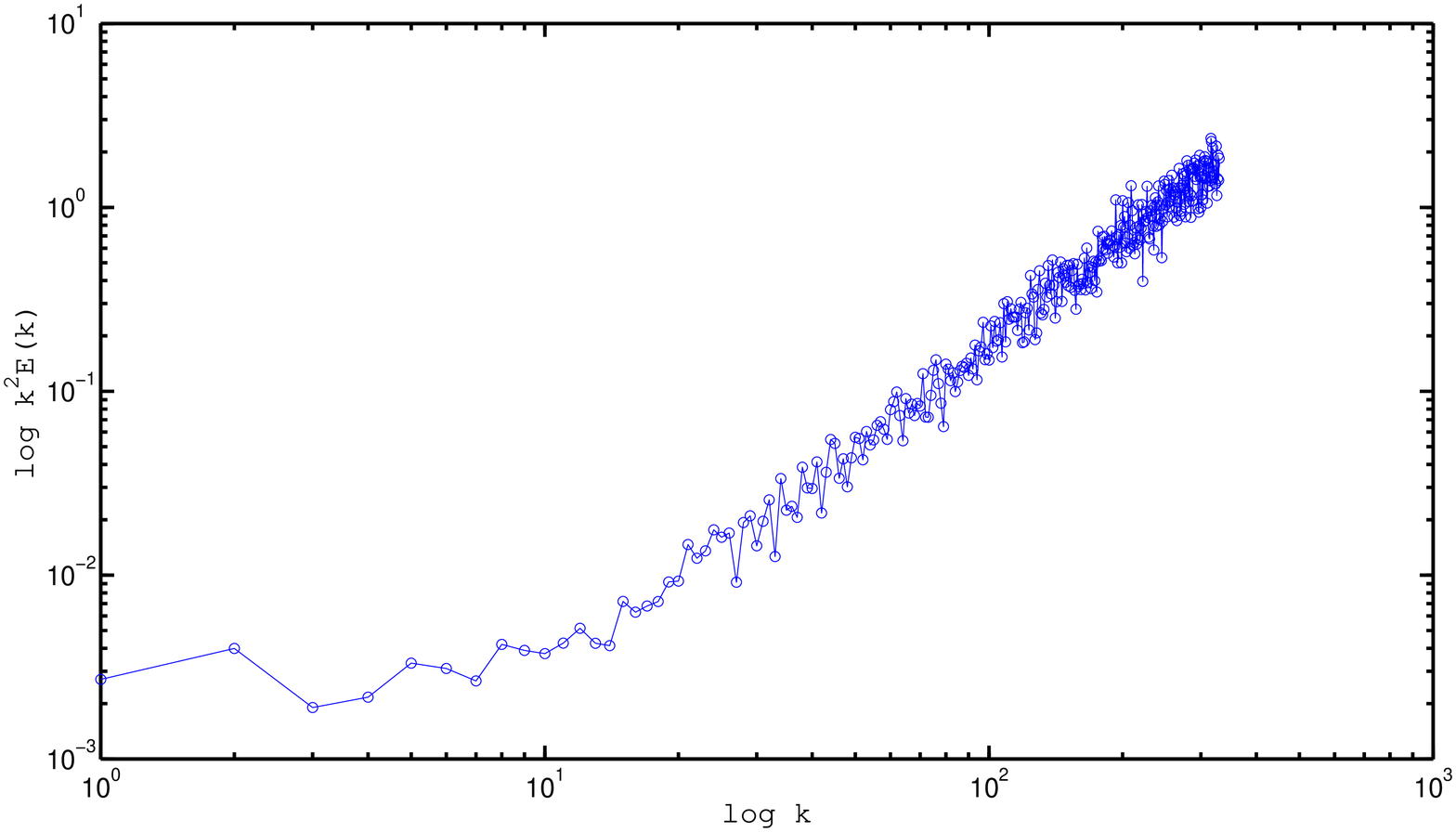}}
\caption{(Color online) (Left) A representative log-log plot of a 
bottleneck in the compensated energy spectrum $k^2E(k)$ of a 
hyperviscous Burgers equation with $\alpha = 4$.
(Right) A representative log-log plot of $k^2E(k)$ versus $k$ 
for $\alpha = 200$ at time $t = 30$. We see clear signatures of 
thermalization at large $k$ (see text).}
\label{bot}
\end{figure}


\subsection{Turbulence with Polymer Additives}

In this Subsection, we present a few results from our numerical
study~\cite{ppdrag} of the analogue drag reduction by polymer
additives in homogeneous, isotropic turbulence. This requires a DNS
of considerably greater complexity than the ones we have described
above. A na\"ive pseudospectral method cannot be used for the FENE-P
model given in Eqs. (18) and (19): the polymer conformation tensor
$\cal{C}$ is symmetric and positive definite; however, in a
practical implementation of the pseudo-spectral method it loses this
property. We have employed a numerical technique that uses a
Cholesky decomposition to overcome this problem; we refer the
reader to Ref.~\cite{ppdrag} for these details. 

Our recent DNS of this model has shown that the natural analogue
drag reduction in decaying, homogeneous, isotropic turbulence is
dissipation reduction; the percentage reduction DR can be defined as
\begin{eqnarray}
{\rm DR}\equiv\left(\frac{\epsilon^{f,m}-\epsilon^{p,m}}
{\epsilon^{f,m}}\right)\times 100;
\label{dragreduction} \end{eqnarray} here the superscripts $f$ and
$p$ stand, respectively, for the fluid without and with polymers and
the superscript $m$ indicates the time $t_m$ at which the cascade is
completed.  The dependence of DR on the polymer concentration
parameter $c$ and the Weissenberg number may be found in
Ref.~\cite{ppdrag}.  Here we show how the addition of polymers
reduces small-scale structures in the turbulent flow: By a
comparison of the isosurfaces of $|\omega|$ in the left (without
polymers) and right (with polymers) panels of Fig. 14, we see that
slender vorticity filaments are suppressed by the polymers; this is
in qualitative agreement with experiments~\cite{hoyt}. The PDFs of
$|\omega|$, with and without polymers (left panel of Fig. 15)
confirm that regions of large vorticity are reduced by polymers. The
right panel of Fig. 15 shows how the polymers modify the energy
spectrum in the dissipation range; this behaviour has been seen in
recent experiments~\cite{bodenpoly}, which study the second-order
structure function that is related simply to the energy spectrum.
For a full discussion of these and related results we refer the
reader to Ref.~\cite{chirag0,ppdrag}.

\begin{figure}[htbp]
\epsfxsize=6cm
\centerline{\epsfbox{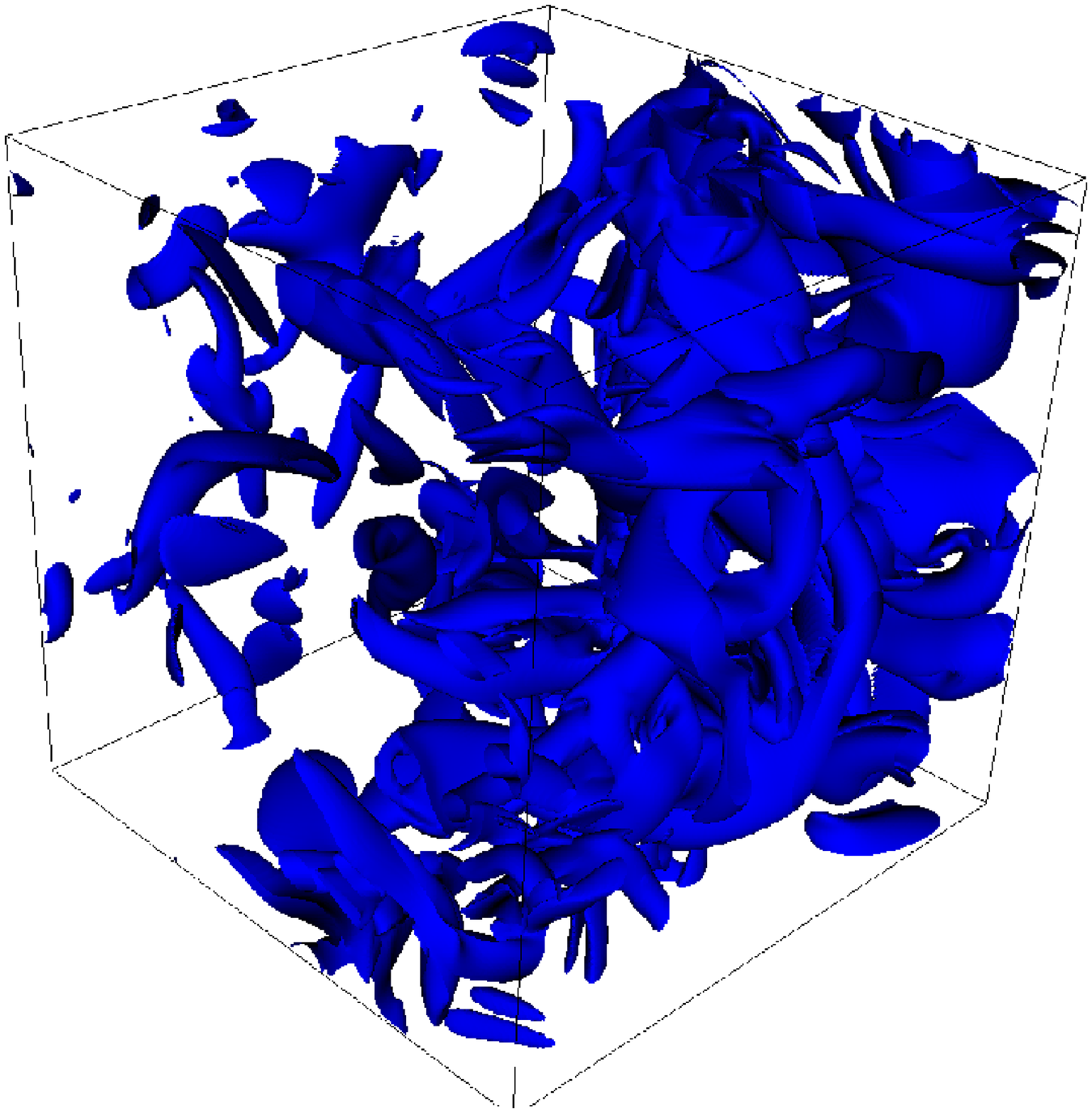},\epsfxsize=6cm
\epsfbox{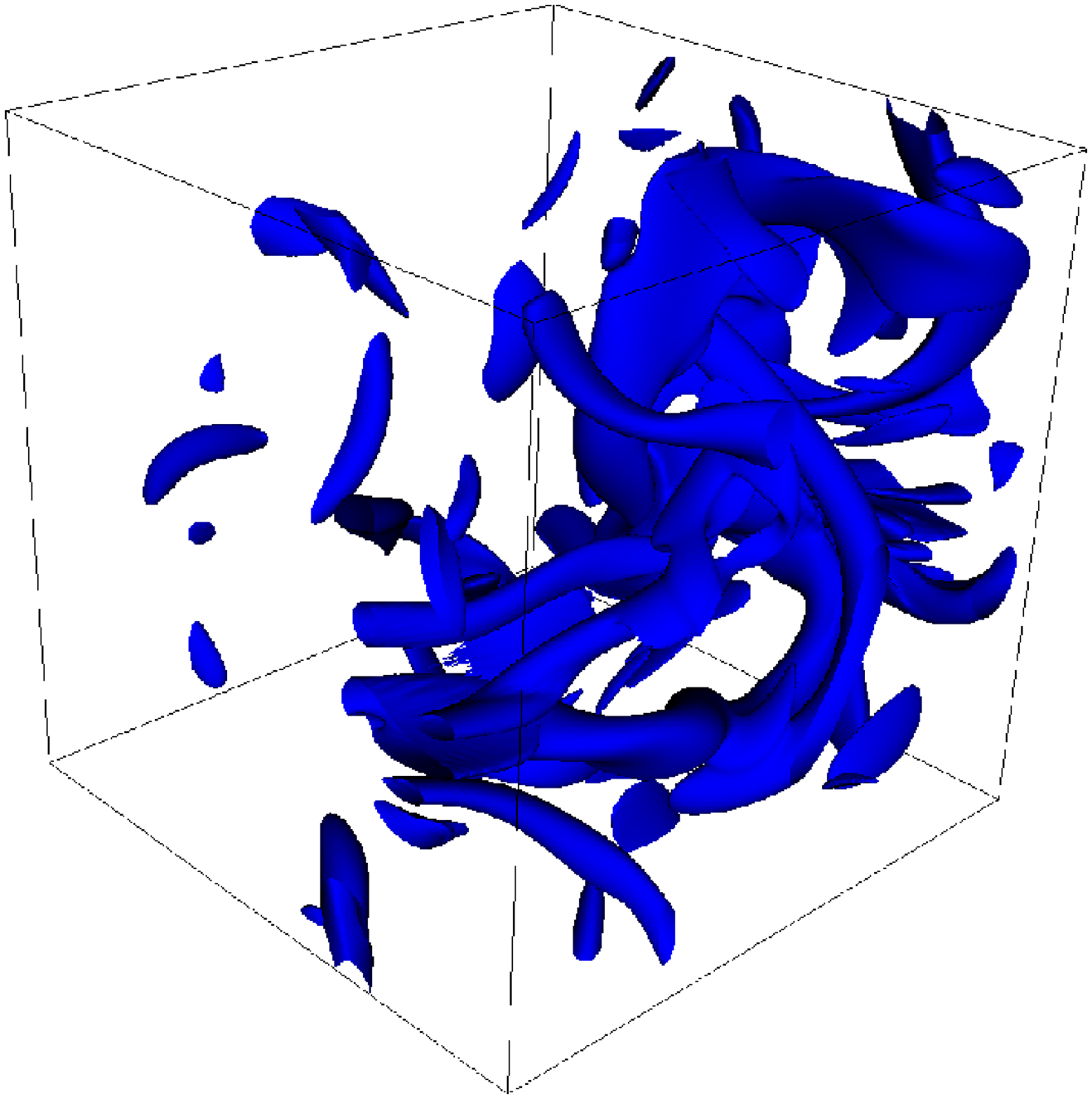}}
\caption{(Color online) Constant-$|\omega|$ isosurfaces for
$|\omega|=\langle{|\omega|}\rangle+\sigma$ at cascade completion
without and (Right) with polymers ($c = 0.4$);
$\langle|\omega|\rangle$ is the mean
and $\sigma$ the standard deviation of $|\omega|$.} 
\label{poly1}
\end{figure}

\begin{figure}[htbp]
\epsfxsize=6cm
\centerline{\epsfbox{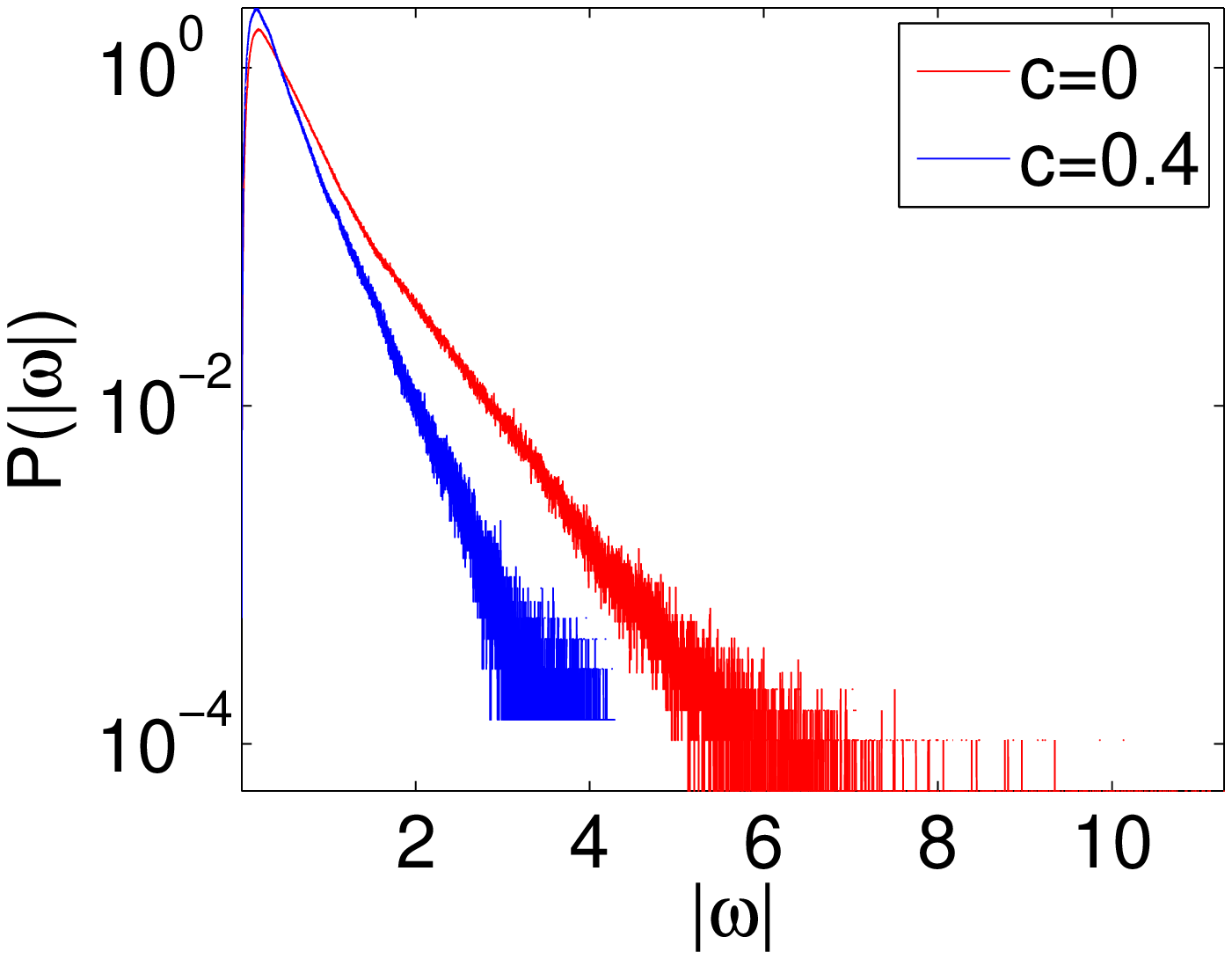},\epsfxsize=6cm
\epsfbox{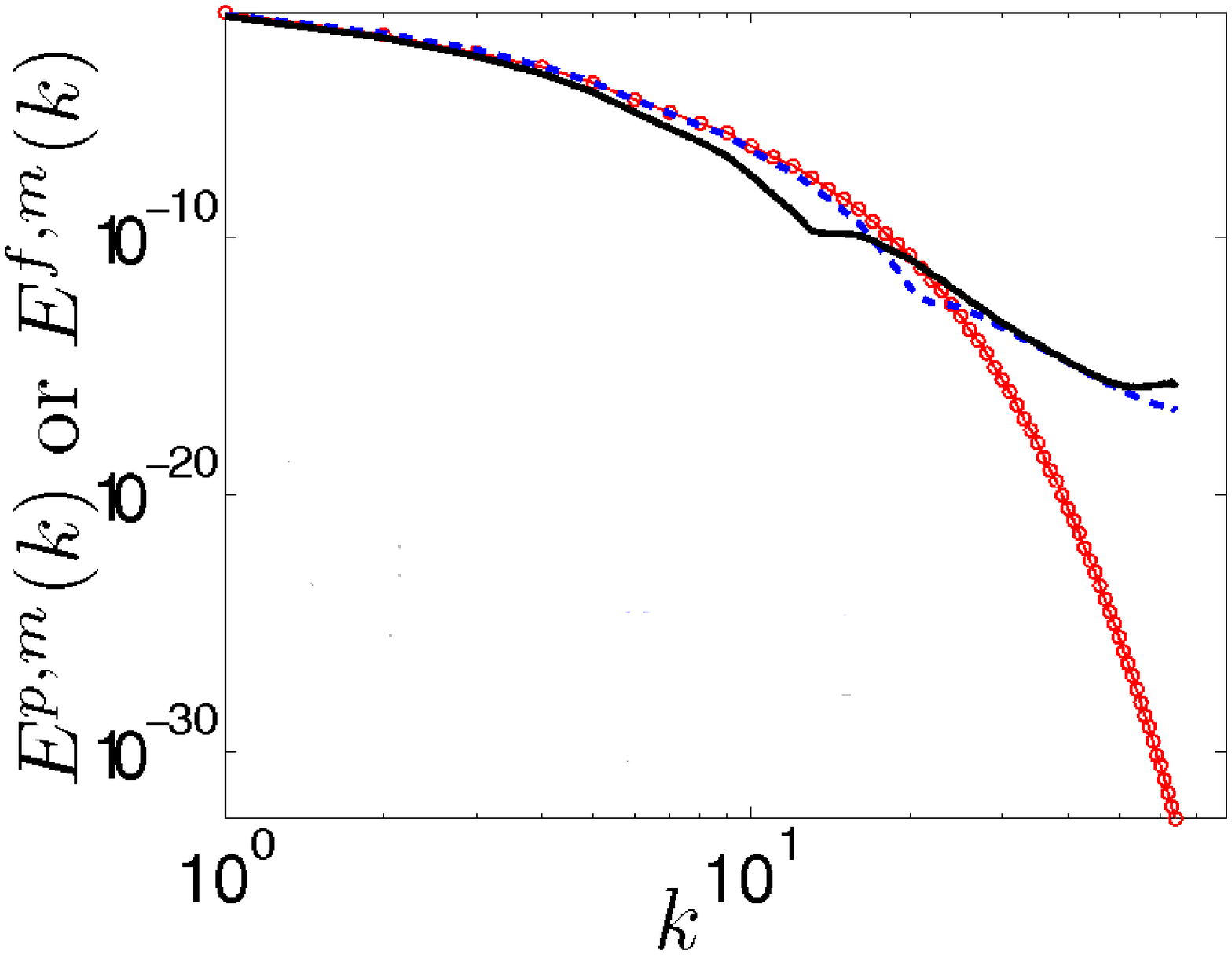}}
\caption{(Color online) (Left) PDF of $\omega$ at cascade 
completion without ($c=0$) and with polymers ($c=0.4$). 
Note that regions of large vorticity are reduced on the addition 
of polymers. (Right) Representative plots of the energy spectra 
$E^{p,m}(k)$ or $E^{f,m}(k)$ versus $k$ 
for $c = 0.1$ (blue dashed line) and 
$c=0.4$ (solid line).} 
\label{poly2}
\end{figure}

\section{Conclusions}

Turbulence provides us with a variety of challenging problems.  We
have tried to give an overview of some of these, especially those
that deal with the statistical properties of turbulence.  The choice
of topics has been influenced, of course, by the areas in which we
have carried out research. For complementary, recent overviews we
refer the reader to Refs.~\cite{ecke,falcosreeni,procsreeni}; 
we hope the other reviews and books that we have cited to 
will provide the reader with further details.

We would like to thank CSIR, DST, and UGC
(India) for support, and SERC (IISc) for computational resources.

\end{document}